\documentclass{article}

\usepackage{microtype}
\usepackage{graphicx}
\usepackage{subcaption}
\usepackage{booktabs}
\usepackage{hyperref}

\usepackage[preprint]{icml2026}
\usepackage{amsmath}
\usepackage{amssymb}
\usepackage{mathtools}
\usepackage{amsthm}
\usepackage{wrapfig}

\graphicspath{{figures/}}
\icmltitlerunning{Who Moved My \texorpdfstring{$Z$}{Z} Boson?}

\begin{document}

\twocolumn[
  \icmltitle{Who Moved My \texorpdfstring{$Z$}{Z} Boson? Revealing Resonance Features Inside a Neural Network}

  \icmlsetsymbol{equal}{*}

  \begin{icmlauthorlist}
    \icmlauthor{Kevin A. Urqu\'ia-Calder\'on}{nbi}
    \icmlauthor{Midori Kato}{nbi}
    \icmlauthor{Inar Timiryasov}{nbi}
    \icmlauthor{Oleg Ruchayskiy}{nbi}
  \end{icmlauthorlist}

  \icmlaffiliation{nbi}{Niels Bohr Institute, University of Copenhagen, Jagtvej 155A, DK-2200, Copenhagen, Denmark}

  \icmlcorrespondingauthor{Kevin A. Urqu\'ia-Calder\'on}{kevin.urquia@nbi.ku.dk}

  \icmlkeywords{interpretability, flow matching, steering vectors, collider physics}

  \vskip 0.3in
]

\printAffiliationsAndNotice{}

\begin{abstract}
Methods used in mechanistic interpretability have revealed human-aligned concepts inside language and image models.
In this paper, we search for physics concepts inside a neural a generative collider physics model trained on real LHC data (\texttt{ShellFlow}). 
We discover features inside \texttt{ShellFlow} that represent resonances, such as the $Z$, $J/\psi$, and $\Upsilon$.
The network was never told which muons come from a resonance, it was only trained to generate the kinematics of the final-state particles. 
We use these features to alter aspects of the $Z$ boson: we change its mass between $\sim80$ and 105~GeV, and removing such a feature from the neural network produces a broader $Z$ peak. In both cases the rest of the invariant mass spectrum remains mostly unchanged.
The feature was built from the gradients of the neural network, and its selectivity was built using pattern recognition methods.
A generative particle physics model trained on unlabeled data can be modified internally to change a physical quantity.
\end{abstract}

\section{Introduction}
\label{sec:intro}

Language and image models learn internal directions that align with human concepts, and adding such a direction to
the activations, a \textit{steering vector}, changes what the model produces in a controlled
way~\citep{mikolov-2013-linguistic,park-2023-linear,turner-2023-actadd,rimsky-2024-caa,zou-2023-repe,templeton-2024-scaling,sofroniew-2026-emotions}.
The same idea has been carried to image diffusion
models~\citep{haas-2024-discovering,gaintseva-2025-casteer,konovalova-2026-shift} and to models of the natural
sciences~\citep{simon-2025-interplm,huang-2025-steering,dimitrievikj-2025-steering,fear-2025-physics}, including
particle physics~\citep{faucett-2021-mapping,wang-2024-jet,bonnet-2026-neutrino}. Generative models are now also trained directly on real collision data from the Large Hadron Collider (LHC). 
\texttt{ShellFlow}~\citep{kato-2026-shellflow} is a Riemannian flow-matching model trained on ATLAS Open Data~\citep{atlas-2025-opendata}. 
It accurately reproduces the kinematics of LHC collisions, and most remarkably, it also replicates inter-particle correlations, such as the invariant mass spectrum of muon pairs, in particular the $J/\psi$,
$\Upsilon$, and $Z$ peaks. 
This raises a question: can a physical quantity, like the mass of a resonance, be steered along a direction in such a model?
Such a direction would let us test whether a generative model trained on collision data has learned the physics of a resonance, and change a single physical parameter of the generated sample without retraining.

In most of the steering literature, the effect of a steering vector is judged by a classifier or a similarity
score~\citep{turner-2023-actadd,rimsky-2024-caa,gaintseva-2025-casteer,konovalova-2026-shift}.
These scores are used because of a major limitation inherent in images or text: there is no rigorous ground truth, so it is impossible to measure how much a quantity changes.
You cannot measure how much a smile has changed in a picture.
A generator of collision events does not have this limitation: its output is a set of four-momenta, on which a physical observable, such as the invariant mass of a muon pair, can be measured.

In this paper, we search for a direction in \texttt{ShellFlow} that is aligned with the concept of the $Z$ boson.
We aim to find a steering vector that changes the mass of the $Z$ boson and leaves the rest of the dimuon spectrum unchanged. 
We find its shape by doing a Taylor expansion of the invariant mass with respect to the activations of the network, and keeping only the linear term. 
We then argue that such steering vectors are the ones that give the largest second moment of the gradients, the derivatives of the invariant mass with respect to the activations.
We use a Fukunaga--Koontz transform (FKT)~\citep{fukunaga-1970-application} to find the vector along which this second moment is the largest for the $Z$, and the smallest for the other parts of the invariant mass spectrum.
We scale and add this vector to the muon tokens and study the generated events.

We use this vector to change the mass of the $Z$ boson without changing much the $J/\psi$ and $\Upsilon$ peaks, or the continuum.
This steering of the mass is predictable: we can select the mass of the $Z$ pole itself. However, the steering also broadens the $Z$ peak.
Furthermore, we test if this is a direction that the network uses to construct its \textit{internal concept} of the $Z$ boson, by ablating the direction in the residual stream, where we find that the model generates a broader $Z$ bosons without this direction.

\section{Previous work}
\label{sec:previous}
\paragraph{Activation steering.}
Early works looking for linear directions that \textit{point} to concepts inside of a model
were~\citep{mikolov-2013-linguistic,alain-2016-probes,kim-2018-tcav}, later formalised as the linear representation
hypothesis~\citep{park-2023-linear}.
Such linear directions were then used to steer the output of a model
in~\citep{subramani-2022-extracting,turner-2023-actadd} with the use of steering vectors.
Steering vectors have been built in the literature in different ways such as:
differences of activations~\citep{turner-2023-actadd,rimsky-2024-caa,arditi-2024-refusal,li-2023-iti},
a PCA of differences of activations~\citep{zou-2023-repe},
sums of attention-head outputs~\citep{todd-2024-function},
sparse-autoencoder features~\citep{templeton-2024-scaling},
and vectors optimised by gradient descent~\citep{subramani-2022-extracting,dunefsky-2025-oneshot}.
Our steering vectors were obtained from a generalized eigenproblem that contrasts classes, on the gradients of the
model's own output. The closest methods to ours are~\citep{beaglehole-2026-universal,wang-2026-general}, which take
eigenvectors of the ``Average Gradient Outer Product'' of a classifier trained on the activations.
Attribution patching~\citep{nanda-2023-attribution,syed-2023-attribution} also uses gradients with respect to the
activations, but to locate the components and circuits responsible for a behaviour, not to build a steering direction.

\paragraph{Steering in diffusion and flow-matching models.}
Various steering analyses have been done on diffusion and flow matching models.
This includes finding steering vectors in audio~\citep{staniszewski-2026-tada} and
video~\citep{hong-2026-lalqr} models. However, by far, the most work has been done in text-to-image work, with
methods using differences of mean activations between contrasting
prompts~\citep{gaintseva-2025-casteer,konovalova-2026-shift,li-2026-semantic}, sparse-autoencoder
features~\citep{surkov-2025-onestep,kim-2025-conceptsteerers}, and, in unconditional image models,
gradients~\citep{haas-2024-discovering,wang-2026-general}.
We steer a Riemannian flow-matching generator of collision events that is conditioned on the particle content of
the reconstructed event, rather than on text like most of the examples highlighted before.
This also provides us a more rigorous approach to what the ground truth is with respect to what we are steering, and can even properly check if the steering respects physical laws.

\paragraph{Interpretability in natural science models.}
A growing body of work has studied interpretability in the natural sciences.
In biology, many studies have found and/or steered interpretable features of protein language
models~\citep{simon-2025-interplm,villegasgarcia-2025-interpreting,huang-2025-steering,parsan-2025-towards}.
In chemistry, steering vectors have been found in chemical language
models~\citep{dimitrievikj-2025-steering}.
In physics, interpretability methods have been applied to models of partial differential
equations~\citep{fear-2025-physics,patil-2026-physsae}, weather~\citep{macmillan-2025-mechanistic}, quantum many-body
states~\citep{qi-2026-nqs}, cosmology and
astronomy~\citep{ocampo-2025-interpretable,ocampo-2026-cmb,bonnetguerrini-2026-realbogus,wu-2025-euclidsae}, and
nuclear structure~\citep{kitouni-2024-neurons}. 
In particle physics, they have been applied to jet
classifiers~\citep{faucett-2021-mapping,wang-2024-jet,rai-2026-dissecting}, to parton
distributions~\citep{kriesten-2024-xai4pdf,bonnetguerrini-2026-shapley}, to the latent space of a jet variational
autoencoder~\citep{collins-2021-wjets}, to jet-tagging transformers~\citep{tiourine-2026-activation}, and for a neutrino-telescope foundation model~\citep{bonnet-2026-neutrino}.
This paper, to our knowledge, presents the first application of steering vectors to a generative particle-physics
model.

\section{Background}
\label{sec:background}

\subsection{Flow matching models}
\label{sec:flowmatching}
Conditional flow matching (CFM)~\citep{lipman-2023-flow} is a machine learning method that generates data.
In it, a network, with parameters $\theta$, outputs a velocity field, $\mathbf{u}_t^\theta$, that transforms an initial distribution
$p_0$ to the data distribution $p_1$. Data are generated by solving the ODE
\begin{equation}
    \label{eq:ode}
    \dot{\mathbf{x}}_t = \mathbf{u}_t^\theta(\mathbf{x}_t)\,, \qquad \mathbf{x}_0 \sim p_0\,,
\end{equation}
from $t = 0$ to $t = 1$. Taking $p_0 = \mathcal{N}(0, \mathbf{I})$, we can build a path between $\mathbf{x}_0 \sim p_0$ and
$\mathbf{x}_1 \sim p_1$,
\begin{equation}
    \label{eq:euclidean_path}
    \begin{aligned}
    \mathbf{x}_t &= (1 - t)\,\mathbf{x}_0 + t\,\mathbf{x}_1\,, \\
    \mathbf{u}_t(\mathbf{x}_t \mid \mathbf{x}_1) &= \mathbf{x}_1 - \mathbf{x}_0 = \frac{\mathbf{x}_1 - \mathbf{x}_t}{1 - t}\,.
    \end{aligned}
\end{equation}
The network is trained on this conditional velocity with the loss
\begin{equation}
    \label{eq:loss_CFM}
    \mathcal{L}_\mathrm{CFM}(\theta) = \mathbb{E}_{t,\,\mathbf{x}_0,\,\mathbf{x}_1}
    \left\| \mathbf{u}_t^\theta(\mathbf{x}_t) - \mathbf{u}_t(\mathbf{x}_t \mid \mathbf{x}_1) \right\|^2\,.
\end{equation}
Data are generated with successive Euler steps, $\mathbf{x}_{t+\Delta t} = \mathbf{x}_t + \Delta t\,\mathbf{u}_t^\theta(\mathbf{x}_t)$, that end in an $\mathbf{x}_1$ that should resemble the data. $\Delta t = 1/N$ for $N$ Euler steps.

Riemannian flow matching~\citep{chen-2024-riemannian} generalizes CFM to a smooth Riemannian manifold $\mathcal{M}$
with metric $g$. The velocity now lies in the tangent space, $\mathbf{u}_t^\theta(\mathbf{x}) \in T_{\mathbf{x}}\mathcal{M}$, the norm in the loss
is taken in the metric, $\|\cdot\|_g$, and the straight line of Eq.~\eqref{eq:euclidean_path} becomes the geodesic,
\begin{equation}
    \label{eq:riemannian_path}
    \begin{aligned}
    \mathbf{x}_t &= \exp_{\mathbf{x}_1}\!\left((1 - t)\log_{\mathbf{x}_1}(\mathbf{x}_0)\right)\,, \\
    \mathbf{u}_t(\mathbf{x}_t \mid \mathbf{x}_1) &= \frac{\log_{\mathbf{x}_t}(\mathbf{x}_1)}{1 - t}\,.
    \end{aligned}
\end{equation}
Data generation proceeds through the same Euler steps, now with the exponential map,
$\mathbf{x}_{t+\Delta t} = \exp_{\mathbf{x}_t}\!\left(\Delta t \,\mathbf{u}_t^\theta(\mathbf{x}_t)\right)$, so that $\mathbf{x}_t$ remains on $\mathcal{M}$ for all steps.

\subsection{ShellFlow}
\label{sec:shellflow}
In a previous work we introduced \texttt{ShellFlow}~\citep{kato-2026-shellflow}, a Riemannian conditional flow-matching transformer that generates
the kinematics of reconstructed collider events, given the particle composition of the
event.

It was trained on unlabelled real data: a mixed dataset of the two-to-four-lepton~\citep{atlas-2025-opendata-2to4lep} and the
1LMET30~\citep{atlas-2025-opendata-1lmet30} skims of the ATLAS Open Data 13~TeV release~\citep{atlas-2025-opendata}.
We trained only on events with, at most, eight particles. In total the dataset has
$\sim$825 million events; the training-validation split is 95/5.

The model has two different manifolds, one for massless particles (electrons, muons, and photons) and another for
massive particles (jets).
The massless manifold is $\mathbb{R}\times S^2$. In it, $S^2$ parametrizes the direction of the momentum,
$\hat{\mathbf{n}} = \mathbf{p}/\lvert\mathbf{p}\rvert$, and $\mathbb{R}$ encodes the energy information through
$y_E = \log(E/E_{\mathrm{ref}})$, with a reference energy $E_{\mathrm{ref}}$ for each particle type.
At $t = 0$, the coordinates are sampled from $p_0$: $y_E \sim \mathcal{N}(0, 1)$ and $\hat{\mathbf{n}} \sim \mathrm{Unif}(S^2)$.
The massive manifold is $\mathbb{R}\times S^2\times\mathbb{R}$: the first $\mathbb{R}$ is the boost rapidity, and the
second the logarithm of the mass (both sampled from a standard normal distribution).
In both, the momentum follows from the on-shell condition $E^2 = p^2 + m^2$, so every generated particle is on its
mass shell.

Each particle of an event is one token. The network has six sequential transformer blocks with a width of 128, with 4 heads, cross-attention with missing
transverse momentum, and a feed-forward layer. The blocks are conditioned on the flow time $t$ by an
adaLN-Zero layer~\citep{peebles-2023-dit}. Moreover, the model is conditioned with a feature-wise linear
modulation~\citep{perez-2018-film} that embeds the particle type, charge, and detector quantities. In total, the model
has $\sim$3 million parameters.

The model has two losses: the usual flow-matching loss, similar to the one in Eq.~\eqref{eq:loss_CFM}, and an auxiliary loss that acts on a different velocity that aims at
reconstructing the invariant masses of two, three, and four objects. The latter does not make the model
agnostic to the structure of an invariant mass. We use the vanilla velocity field to generate data with
200 Euler steps, given the composition of the event.

\subsection{Muon pair production at the LHC}
\begin{figure}[t]
\centering
\includegraphics[width=0.8\columnwidth]{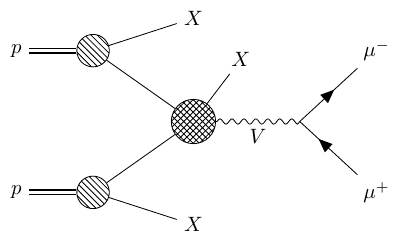}
\caption{Feynman diagram of the production of a $\mu^-\mu^+$ pair, through an intermediate vector boson
$V = J/\psi, \Upsilon, Z, \gamma$. $X$ denotes the proton remnants and any additional radiation.}
\label{fig:feynman}
\end{figure}

Collisions at the LHC that result in a pair of muons are very important for analyses done at the LHC collaborations as
a whole, as these are used to calibrate the measurement of energy and momentum of muons~\citep{atlas-2023-studies}.
In particular, the processes through a resonant vector particle, like the $J/\psi$, $\Upsilon$, or $Z$, are used,
since their masses are known precisely and fix the momentum scale of the muons.
A typical process that produces a pair of muons at the LHC is shown in Fig.~\ref{fig:feynman}.

A resonance occurs when the partons, the constituent particles of the protons, have a centre-of-mass energy equal to
the mass of the resonant particle. The outgoing muons, with momenta $p_\pm = (E_\pm, \mathbf{p}_\pm)$, then have an
invariant mass of
\begin{equation}
    \label{eq:invariant_mass}
    m_{\mu\mu}^2 \simeq 2\,\left(E_{+}E_{-} - \mathbf{p}_+\cdot\mathbf{p}_-\right)\,,
\end{equation}
where the mass of the muons is neglected. At resonance, the invariant mass of the muon pair is around the mass of the
resonant particle ($m_{J/\psi} = 3.10$~GeV, $m_{\Upsilon} = 9.46$~GeV, $m_Z = 91.19$~GeV~\citep{pdg-2026-review}),
and each appears as a narrow peak on top of the continuum (see Fig.~\ref{fig:spectrum}, grey).

In terms of the \texttt{ShellFlow} parameters, $E = E_{\mathrm{ref}}\,e^{y_E}$ and $\mathbf{p} = E\,\hat{\mathbf{n}}$,
it is more convenient to express Eq.~\eqref{eq:invariant_mass} in terms of $\log m_{\mu\mu}$,
\begin{equation}
    \label{eq:log_mass}
    \begin{split}
    \log m_{\mu\mu} &= \log E_{\mathrm{ref}} + \frac{1}{2}\left(y_{E_+} + y_{E_-}\right) \\
                    &\quad + \frac{1}{2}\log\left[2\,(1 - \hat{\mathbf{n}}_+\cdot\hat{\mathbf{n}}_-)\right]\,,
    \end{split}
\end{equation}
where $E_{\mathrm{ref}}$ is the same for both muons.

\label{sec:physics}

\section{Method}
\label{sec:method}
\begin{figure*}[t]
\centering
\includegraphics[width = \textwidth]{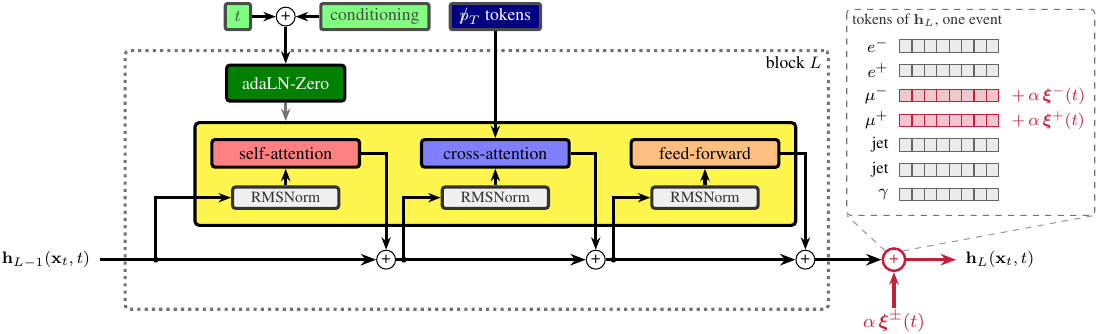}
\caption{One transformer block of \texttt{ShellFlow}, block $L$ (dotted frame). The steering vector is added to the
residual stream at the output of the block. Inset: the output tokens of one event; only the $\mu^-$ and $\mu^+$
tokens receive the steering vector, every other token is left unchanged.}
\label{fig:block}
\end{figure*}

Our main objective in this paper is to identify which directions in the residual stream of \texttt{ShellFlow}
change a physical quantity in a controlled way.
In Sec.~\ref{sec:physics}, we summarized the importance of known resonances to the study of collider physics.
Here we aim to find a vector that only affects the muon pairs with an invariant mass close to a specific resonance,
and that can alter one of its physical observables, in particular its mass.

We will do this by adding a steering vector to the residual stream, of the form
\begin{equation}
    \mathbf{h}_L \to \mathbf{h}_L + \alpha\,\boldsymbol{\xi}\,,
    \label{eq:edit}
\end{equation}
where $\mathbf{h}_L$ is the residual stream after block $L$, $\boldsymbol{\xi}$ is the steering vector, and $\alpha$ is the steering
coefficient.

In the rest of this section, we outline a method to obtain such a steering vector.

\subsection{Steering vectors and gradients}
\label{sec:gradients}
We shall now write Eq.~\eqref{eq:edit} with its explicit indices and dependences
\begin{equation}
    \label{eq:edit-full}
    \mathbf{h}_{L}^{\mu^\pm}(\mathbf{x}_t, t) \to \mathbf{h}_{L}^{\mu^\pm}(\mathbf{x}_t, t) + \alpha\,\boldsymbol{\xi}^{\pm}(t)\,,
\end{equation}
where $\mathbf{h}_{L}^{\mu^\pm} \in \mathbb{R}^{128}$ is the residual stream at block $L \in \{0, \dots, 5\}$, at flow time $t$ and state of the event $\mathbf{x}_t$, on the token of the $\mu^+$ or $\mu^-$ of the event.
The steering vector $\boldsymbol{\xi}^{\pm}$ is added to the token of $\mu^{\pm}$, and it also depends on flow time.
We only add the vector to the slot of the muon tokens because we are looking for directions that only affect the shape of the invariant mass of muon pairs, but it can be easily applied to any other particle type.

The steering vector is applied at the output of the block, after its feed-forward layer has been added
(see Fig.~\ref{fig:block}).

The response of the mass to the edit follows from a Taylor expansion of $\log m_{\mu\mu}$ in the added vectors
(the derivation is in App.~\ref{app:derivation}):
\begin{equation}
\label{eq:linresp}
\Delta(\log m_{\mu\mu,p})
   = \alpha \sum_{k=0}^{N-1} \mathbf{G}_p(t_k)^\top\boldsymbol{\xi}(t_k) + \mathcal{O}(\alpha^2)\,.
\end{equation}
The sum runs over the $N = 200$ Euler steps of the sampler, at flow times $t_k = k/N$, and $\Delta\log m_{\mu\mu,p}$ is the
change of the final $\log m_{\mu\mu}$ of pair $p$ with respect to the generation without the edit. $\mathbf{G}_p(t_k)$ and $\boldsymbol{\xi}(t_k)$ are concatenations over the two muon tokens:
\begin{align}
\mathbf{G}_p(t_k) &= \left[\frac{\mathrm{D}\log m_{\mu\mu}}{\mathrm{D}\mathbf{h}_L^{\mu^-}(t_k)}\,,\;
                        \frac{\mathrm{D}\log m_{\mu\mu}}{\mathrm{D}\mathbf{h}_L^{\mu^+}(t_k)}\right] \in \mathbb{R}^{256}\,,
                        \label{eq:read-gradient}\\
\boldsymbol{\xi}(t_k) &= \left[\boldsymbol{\xi}^{-}(t_k)\,,\; \boldsymbol{\xi}^{+}(t_k)\right] \in \mathbb{R}^{256}\,,
                        \label{eq:xi-concat}
\end{align}
where we call $\mathbf{G}_p(t_k)$ the gradient of pair $p$ at step $k$, and $\mathrm{D}$ is the total derivative.

This tells us that, for a steering vector of fixed length, the shift is largest when it is aligned with the gradients.
Moreover, if we want the shift to be localized in a specific region of the invariant mass spectrum, such as near a resonance, the steering vector must be aligned with the gradients of the pairs in that region, and as little as possible with the gradients of all other pairs.\footnote{From Eq.~\eqref{eq:log_mass}, a shift $\log m_{\mu\mu} \to \log(m_{\mu\mu} e^\beta)$ can be achieved by $y_{E_\pm} \to y_{E_\pm} + \beta$. This is trivial to implement at the output of the network. 
However, it is not localized, it rescales the energy of every muon, not only of those in pairs near a specific resonance. 
Moreover, it does not tell us anything about the inner workings of \texttt{ShellFlow}.}

\subsection{Discriminating directions}
\label{sec:directions}
Our task now is to find the vectors, $\boldsymbol{\xi}(t_k)$, that are aligned with the gradients of a specific resonance and as little as possible with other gradients.
We define the \textit{resonance classes} in Table~\ref{tab:classes}, where $m_{\mu\mu}$ is the invariant mass at generation without any
intervention.

We can now pose this as an optimization problem. Let us consider one of the classes in Table~\ref{tab:classes} as
$\mathcal{C}$, with $n_{\mathcal{C}}$ pairs, and group the rest of the classes into $\mathcal{C}^\prime$.
From Eq.~\eqref{eq:linresp}, the response of pair $p$ to a steering vector at step $t_k$ is
\begin{equation}
\alpha\,\mathbf{G}_p(t_k)^{\top}\boldsymbol{\xi}(t_k)
  = \alpha\,\lVert\mathbf{G}_p(t_k)\rVert\,\hat{\mathbf{G}}_p(t_k)^{\top}\boldsymbol{\xi}(t_k)\,,
\label{eq:pair-response}
\end{equation}
with the unit gradient $\hat{\mathbf{G}}_p = \mathbf{G}_p/\lVert\mathbf{G}_p\rVert$.\footnote{We also tried the $\mathbf{G}_p$ themselves, rather than the unit gradients. But their norms span about four
orders of magnitude, and a few pairs dominated the moments.}
We measure the response of a class by its average of the squared alignment,
$\frac{1}{n_{\mathcal{C}}}\sum_{p\in\mathcal{C}}\boldsymbol{\xi}^{\top}\hat{\mathbf{G}}_p\hat{\mathbf{G}}_p^{\top}\boldsymbol{\xi}
= \boldsymbol{\xi}^{\top}\boldsymbol{\Sigma}_{\mathcal{C}}\,\boldsymbol{\xi}$, where
\begin{equation}
\begin{aligned}
\boldsymbol{\Sigma}_{\mathcal{C}}(t_k) &= \frac{1}{n_{\mathcal{C}}}\sum_{p\in\mathcal{C}}
  \hat{\mathbf{G}}_p(t_k)\,\hat{\mathbf{G}}_p(t_k)^{\top} \in \mathbb{R}^{256\times256}\,,\\
\boldsymbol{\Sigma}_{\mathcal{C}^\prime}(t_k) &= \sum_{\mathcal{D}\neq\mathcal{C}}\frac{1}{n_{\mathcal{D}}}\sum_{p\in\mathcal{D}}
  \hat{\mathbf{G}}_p(t_k)\,\hat{\mathbf{G}}_p(t_k)^{\top}\,,
\end{aligned}
\label{eq:moments}
\end{equation}
so every class enters with the same weight, whatever its number of pairs.
We choose the squared response to ensure that every pair of $\mathcal{C}^\prime$ is small. Choosing the mean would
allow the mean response to move up and down and cancel.

Our objective is to maximize the response of $\mathcal{C}$ while minimizing the response of $\mathcal{C}^\prime$, or,
in other words:
\begin{equation}
\boldsymbol{\xi}_{\mathcal{C}} = \operatorname*{arg\,max}_{\boldsymbol{\xi}}\,
  \frac{\boldsymbol{\xi}^{\top}\boldsymbol{\Sigma}_{\mathcal{C}}\,\boldsymbol{\xi}}
       {\boldsymbol{\xi}^{\top}\left(\boldsymbol{\Sigma}_{\mathcal{C}} + \boldsymbol{\Sigma}_{\mathcal{C}^\prime}\right)\boldsymbol{\xi}}\,.
\label{eq:fk-criterion}
\end{equation}
This is the Fukunaga--Koontz criterion~\citep{fukunaga-1970-application}\footnote{The same
method has been used in different areas of research, and usually goes by different names. Uses go from
neuroscience~\citep{blankertz-2008-optimizing}, to cosmology~\citep{tegmark-1997-karhunen}, and climate
science~\citep{delsole-2008-predictable}, among other areas of research.}.
This is equivalent to the optimization of
$\boldsymbol{\xi}^{\top}\boldsymbol{\Sigma}_{\mathcal{C}}\,\boldsymbol{\xi}/\boldsymbol{\xi}^{\top}\boldsymbol{\Sigma}_{\mathcal{C}^\prime}\,\boldsymbol{\xi}$.
We chose the fraction in Eq.~\eqref{eq:fk-criterion} because it lies in $[0,1]$.

Eq.~\eqref{eq:fk-criterion} is a generalized Rayleigh quotient, whose maximum solves the generalized eigenproblem
\begin{equation}
\boldsymbol{\Sigma}_{\mathcal{C}}\,\boldsymbol{\xi}_{\mathcal{C}} = \lambda_{\mathcal{C}}\left(\boldsymbol{\Sigma}_{\mathcal{C}} + \boldsymbol{\Sigma}_{\mathcal{C}^\prime}\right)\boldsymbol{\xi}_{\mathcal{C}}\,.
\label{eq:gen-eigen}
\end{equation}
We solve it by making a Fukunaga--Koontz transformation (FK transformation), that goes as follows (we omit the
dependence on $t_k$ for convenience). We first find the spectral decomposition of the sum of both
$\boldsymbol{\Sigma}$ matrices,
\begin{equation}
\boldsymbol{\Sigma}_{\mathcal{C}\mathcal{C}^\prime} \equiv \boldsymbol{\Sigma}_{\mathcal{C}}
  + \boldsymbol{\Sigma}_{\mathcal{C}^\prime} = \mathbf{U}\,\boldsymbol{\sigma}\,\mathbf{U}^{\top}\,,
\label{eq:fk-sum}
\end{equation}
where $\boldsymbol{\sigma}$ is a diagonal matrix with the eigenvalues of $\boldsymbol{\Sigma}_{\mathcal{C}\mathcal{C}^\prime}$,
all positive, and $\mathbf{U}$ is an orthogonal matrix whose columns are the orthonormal eigenvectors of
$\boldsymbol{\Sigma}_{\mathcal{C}\mathcal{C}^\prime}$. We then transform the gradients,
$\hat{\mathbf{G}}_p \to \mathbf{M}_{\mathrm{FK}}^{\top}\hat{\mathbf{G}}_p$, by the matrix
$\mathbf{M}_{\mathrm{FK}} = \mathbf{U}\,\boldsymbol{\sigma}^{-1/2}$, such that
\begin{equation}
\tilde{\boldsymbol{\Sigma}}_{\mathcal{C}} = \mathbf{M}_{\mathrm{FK}}^{\top}\boldsymbol{\Sigma}_{\mathcal{C}}\,\mathbf{M}_{\mathrm{FK}}\,,\qquad
\tilde{\boldsymbol{\Sigma}}_{\mathcal{C}^\prime} = \mathbf{M}_{\mathrm{FK}}^{\top}\boldsymbol{\Sigma}_{\mathcal{C}^\prime}\,\mathbf{M}_{\mathrm{FK}}\,;
\label{eq:fk-transform}
\end{equation}
then, it immediately follows that $\tilde{\boldsymbol{\Sigma}}_{\mathcal{C}} + \tilde{\boldsymbol{\Sigma}}_{\mathcal{C}^\prime}
= \mathbf{I}$. Now, given an eigenvector $\tilde{\boldsymbol{\xi}}_{\mathcal{C}}$ of $\tilde{\boldsymbol{\Sigma}}_{\mathcal{C}}$
with eigenvalue $\lambda_{\mathcal{C}}$, we get that
\begin{equation}
\tilde{\boldsymbol{\Sigma}}_{\mathcal{C}}\,\tilde{\boldsymbol{\xi}}_{\mathcal{C}} = \lambda_{\mathcal{C}}\,\tilde{\boldsymbol{\xi}}_{\mathcal{C}}\,,\qquad
\tilde{\boldsymbol{\Sigma}}_{\mathcal{C}^\prime}\,\tilde{\boldsymbol{\xi}}_{\mathcal{C}} = (1-\lambda_{\mathcal{C}})\,\tilde{\boldsymbol{\xi}}_{\mathcal{C}}\,.
\label{eq:eigen}
\end{equation}
Eq.~\eqref{eq:eigen} tells us that both FK-transformed matrices share the same eigenbasis, but with different eigenvalues.
So, if $\tilde{\boldsymbol{\Sigma}}_{\mathcal{C}}$ has an eigenvalue $\lambda_{\mathcal{C}}$,
$\tilde{\boldsymbol{\Sigma}}_{\mathcal{C}^\prime}$ has an eigenvalue $1-\lambda_{\mathcal{C}}$, with $0\leq\lambda_{\mathcal{C}}\leq 1$. This means that
the most dominant eigenvectors in one class will be the least dominant in the other class. The eigenvalue $\lambda_{\mathcal{C}}$ is
the value of Eq.~\eqref{eq:fk-criterion}.

We then take the eigenvector of the largest eigenvalue, $\lambda_{\mathcal{C}}(t_k)$, and bring it back to the coordinates of the
residual stream,
$\boldsymbol{\xi}_{\mathcal{C}}(t_k) = \mathbf{M}_{\mathrm{FK}}\tilde{\boldsymbol{\xi}}_{\mathcal{C}}/\left\lVert\mathbf{M}_{\mathrm{FK}}\tilde{\boldsymbol{\xi}}_{\mathcal{C}}\right\rVert$,
as the direction of the steering vector that shifts the invariant mass of the two muons. We set the sign of the
steering vector by choosing the sign where $\bar{\mathbf{G}}_{\mathcal{C}}(t_k)^{\top}\boldsymbol{\xi}_{\mathcal{C}}(t_k) > 0$, where
$\bar{\mathbf{G}}_{\mathcal{C}}(t_k)$ is the average of $\mathbf{G}_p(t_k)$.
The gradients were computed using a single backward pass, done at all Euler steps. This is the same way gradients were
computed in~\citet{nanda-2023-attribution,syed-2023-attribution}.

\bigskip\noindent
We end this section with other attempts at obtaining steering vectors.
We first built them from the activations of pairs at specific resonances, with a difference of means and a linear probe, and tested whether steering with them, or ablating them, would produce more or fewer resonant events.
These vectors changed the generated spectrum, but not in a way that could be physically interpreted.
We tested two alternative methods to the criterion in Eq.~\eqref{eq:fk-criterion}. We compare them in
App.~\ref{app:different_methods}.

\begin{table}[t]
\caption{The selection criteria of our classes of muon pairs, defined from the invariant mass of the generated pair of one seed.
All events have exactly one pair of opposite-sign muons. 
The $\Upsilon$ class contains all three $\Upsilon$ resonances,
as \texttt{ShellFlow} cannot properly generate three different resonances.}
\label{tab:classes}
\centering
\small
\begin{tabular}{llr}
\toprule
Class & $m_{\mu\mu}$ [GeV] & Events used for the gradients \\
\midrule
$Z$          & $[81.19,\,101.19]$ & 181,987 \\
$J/\psi$   & $[2.9,\,3.3]$      & 224,901 \\
$\Upsilon$ & $[9.2,\,10.6]$     & 50,528 \\
rest       & elsewhere above 0.3 & 542,584 \\
\midrule
total      &                    & 1,000,000 \\
\bottomrule
\end{tabular}
\end{table}

\section{Experiments}
\label{sec:experiments}
\begin{figure*}[t]
\centering
\includegraphics[width = \textwidth]{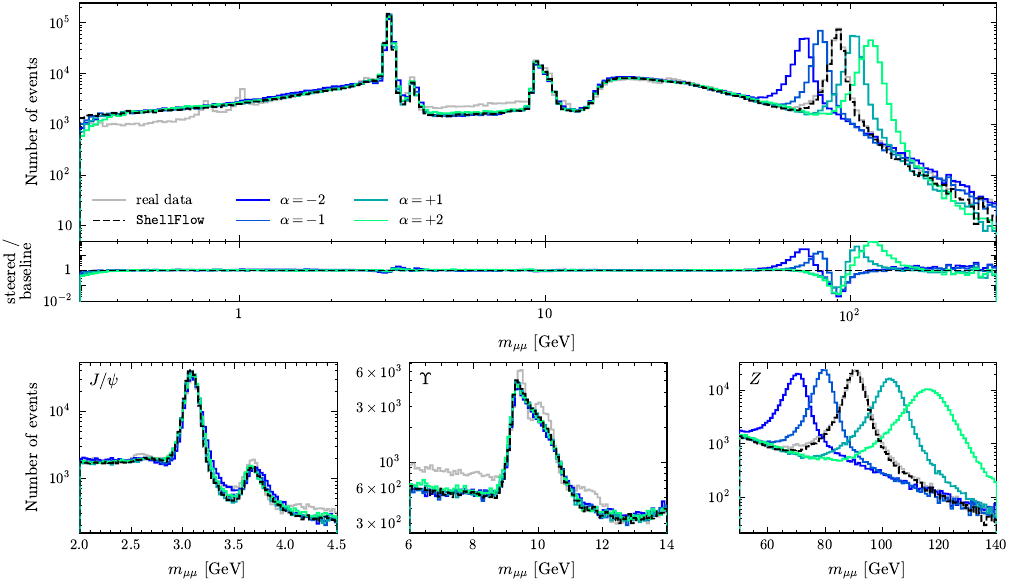}
\caption{The steering of the $Z$ boson mass. Top: dimuon invariant mass of one million muon pairs, compared with real data (grey), baseline generation of \texttt{ShellFlow} (black, dashed), and the results from adding a steering vector at block 1 with $\alpha = \pm 1, \pm 2$ (blue to green). Middle: steered over baseline. Bottom: zoom-in over the different peaks: $J/\psi$, $\Upsilon$, and $Z$ bosons.}
\label{fig:spectrum}
\end{figure*}

In this Section we will now show the results of the \textit{steered generation}. 
We will present how the steering of the $Z$ affects the invariant mass spectrum and the rest of the particle kinematics. We will also show how fitted physical parameters from a phenomenologically inspired model change with $\alpha$. 
Then, we shall present results of the steering vectors themselves, such as their alignment with gradients, and their evolution with $t$. 
Results for the $J/\psi$ and $\Upsilon$ resonances are presented in App.~\ref{app:quarkonia}, more details on the FKT eigenvalues in App.~\ref{app:time}, and the same steering on two other networks in App.~\ref{app:othernetworks}.

To define the selectivity of the steering, we define the following metric, which measures how much the invariant mass spectrum changes $m_{\mu\mu} < 60$~GeV
\begin{equation}
    \mathcal{R}_{60} = \frac{1}{n_b}\sum_{b}\left|\frac{N_b}{N_b^{\mathrm{ShellFlow}}} - 1\right|\,,
    \label{eq:r60}
\end{equation}
where $N_b$ and $N_b^{\mathrm{ShellFlow}}$ are the numbers of muon pairs in bin $b$ of the modified and the baseline generation.
We only compute this metric for bins with $m_{\mu\mu} < 60$~GeV to not include differences from the steering of the $Z$ boson.

\begin{figure*}[t]
\centering
\includegraphics[width = \textwidth]{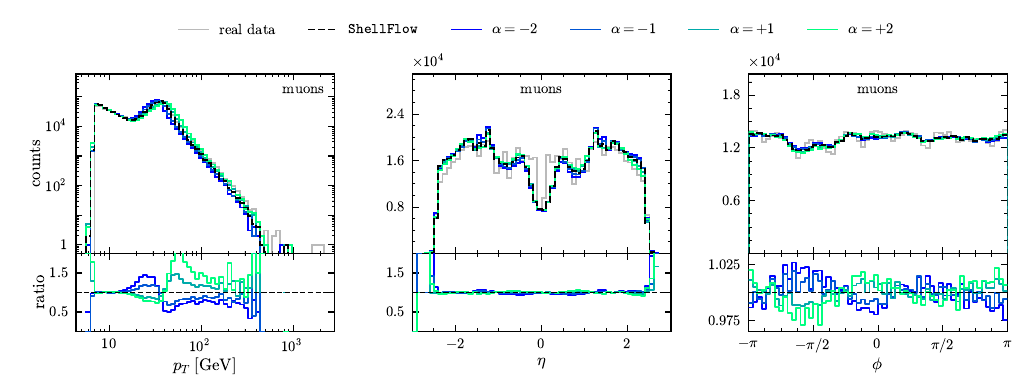}
\caption{Muon kinematics under $Z$ steering. 1,000,000 validation events of every final state, the $Z$ steering vector at
block~1 added to every muon: transverse momentum, pseudorapidity, and azimuth of the muons. Real data (grey), baseline
generation (black, dashed), and steered with $\alpha = -2, -1, +1, +2$ (blue to green); ratio panels: steered
over baseline, drawn in every bin with baseline muons, each panel on its own scale. Every other particle type is in
App.~\ref{app:allparticles}.}
\label{fig:kinematics}
\end{figure*}

\subsection{Steering effects on generated events}
\label{sec:steering_effects}
At every Euler step, we add the steering vector to the muon tokens as in Eq.~\eqref{eq:edit-full} (see also
Fig.~\ref{fig:block}).
In this section, we present results only for block $L = 1$; the reasons for this choice will become evident in the
next section.
We scan $\alpha \in [-3, 3]$ in steps of $0.5$.
Each generation uses 1,000,000 muon pairs from validation events, a different set of events from the one used to
obtain the gradients of the classes in Table~\ref{tab:classes}.
We repeat every generation with three noise seeds and average the results.
Within a seed, the steered and the baseline generations start from the same noise, so every muon pair is compared with itself.

Figure~\ref{fig:spectrum} shows the effects of the $Z$ steering vector on the dimuon invariant mass for
$\alpha = \pm 1, \pm 2$.
The $Z$ peak moves between $\sim$70~GeV and $\sim$115~GeV for the values of $\alpha$ shown in the plot.
The rest of the spectrum changes much less.
We find that $\mathcal{R}_{60} < 0.04$ for $|\alpha| \le 1$.
Most of this change sits around the $J/\psi$ peak.
$\mathcal{R}_{60}$ grows with $|\alpha|$, to 0.112 for $\alpha = -2$, and 0.071 for $\alpha = +2$.
The steering is therefore not symmetric in $\alpha$: the $Z$ peak moves by a similar amount in both directions, but
negative values of $\alpha$ change the continuum more than positive values of the same size.

In App.~\ref{app:lineshapefit} we develop a phenomenologically inspired $Z$ lineshape for hadron colliders.
It is built from the tree-level amplitude at quark level, which we use to derive the proton-level differential cross
section; this cross section is modified to include final-state radiation and convolved with a Gaussian that accounts
for the detector resolution.
We use it to measure how physical parameters, such as the pole mass $M$ and the decay width $\Gamma$, change with
$\alpha$.
We fit further parameters, the detector resolution, a power-law correction to the parton luminosity, and a
normalisation, whose details are in App.~\ref{app:fitdetails}.

\begin{figure}[t]
\centering
\includegraphics[width = 0.45\textwidth]{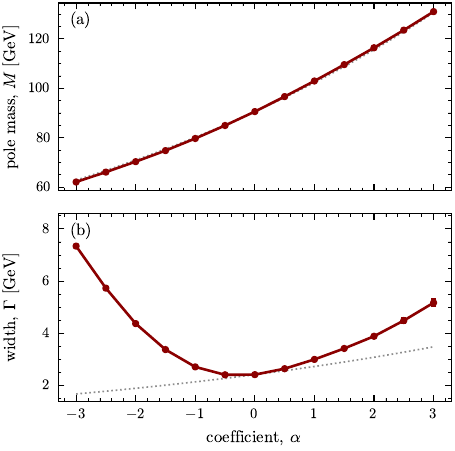}
\caption{Fitted pole and width for different $\alpha$ as steered in block $L=1$, from the lineshape fit of
App.~\ref{app:lineshapefit}. (a) Pole mass $M$. (b) Width $\Gamma$. 
The point at $\alpha=0$ is the baseline fit. 
Dotted: the
baseline value times $\exp\left(\alpha\sum_{k}\bar{\mathbf{G}}_{Z}(t_k)^\top\boldsymbol{\xi}_{Z}(t_k)\right)$,
with the sum equal to 0.122, as predicted by Eq.~\eqref{eq:linresp}. 
Error bars: standard deviation over three seeds of the sampler.}
\label{fig:ladder}
\end{figure}

The fit from the baseline generation of \texttt{ShellFlow} gives $M = 90.57$~GeV and $\Gamma = 2.41$~GeV (real data:
90.98~GeV and 2.87~GeV).
The dependence of $M$ and $\Gamma$ on $\alpha$ is shown in Fig.~\ref{fig:ladder}.
For $-3 \le \alpha \le 3$ the fitted pole moves between 62~GeV and 131~GeV, within 0.9~GeV of the first-order
prediction of Eq.~\eqref{eq:linresp}.
Since the steering vector moves $\log m_{\mu\mu}$, a pure shift of the pole would scale the width by the same factor
as $M$; instead, the peak broadens with $|\alpha|$, which increases $\Gamma$ (in our fit, $\Gamma$ and the detector
resolution are strongly correlated).
For the baseline generation the deviance per degree of freedom is about 2; it stays between 1.8 and 4.3 for
$\alpha \ge -1$, but degrades for more negative values (see App.~\ref{app:fitdetails}).

The steering vector is written on the muon tokens, so the muon kinematics must change with the pair mass
(Fig.~\ref{fig:kinematics}).
On 1,000,000 validation events of every final state, with the steering vector added to every muon, the muon
transverse momentum, $p_T$, changes by up to 32 and 49\% per bin at $\alpha = -1$ and $+1$, around the $Z$ peak near
$M/2$, as expected.
The pseudorapidity, $\eta$, changes by at most 5.4\% per bin, and the azimuth angle, $\phi$, by at most 2\%.
The steering vector could also affect the other particles indirectly, but these effects are small
(App.~\ref{app:allparticles}).

As a control, we add a random vector at $\alpha = \pm 1, \pm 2$. 
This random vector has the same length as the steering vector, is drawn independently for each step, is different for the two muon tokens, and is the same for every event.
The random vector does not steer the $Z$ boson: its pole stays at 89 -- 92~GeV, while our steering vector does move it.
The random vector also disturbs the rest of the spectrum more than the steering vector does, and broadens
the $Z$ peak more.

\subsection{Shape and evolution of the steering vectors}
\label{sec:steering_shape}

\begin{figure}[t]
\centering
\includegraphics[width = \columnwidth]{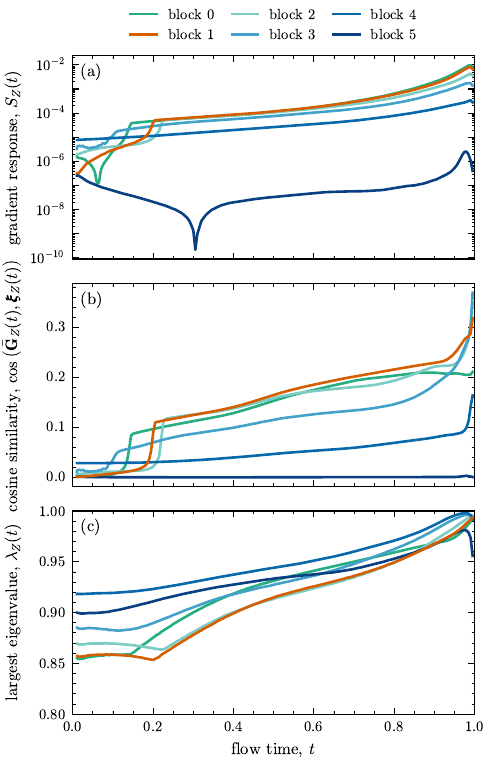}
\caption{The $Z$ steering quality features as a function of flow time $t$. (a) Gradient response of the $Z$,
$S_{Z}(t_k)$, (b) the cosine similarity between the class-mean gradient
of the $Z$ and the steering vector, $\cos\left(\bar{\mathbf{G}}_{Z}(t_k), \boldsymbol{\xi}_{Z}(t_k)\right)$, and (c) the
largest eigenvalue of Eq.~\eqref{eq:eigen} for the $Z$, $\lambda_{Z}(t_k)$.}
\label{fig:quality_steering}
\end{figure}

We can determine how much the $Z$ resonance will change for a given steering vector from Eq.~\eqref{eq:linresp}.
For a single pair $p$, we can see that
\begin{equation}
    m_{\mu\mu,p}(\alpha) \simeq m_{\mu\mu,p}(0)\,\exp\left(\alpha \sum_k \mathbf{G}_p(t_k)^\top\boldsymbol{\xi}_{\mathcal{C}}(t_k)\right)\,,
    \label{eq:pair_steering}
\end{equation}
where we neglected non-linear contributions of $\alpha$, $m_{\mu\mu,p}(0)$ is the invariant mass of the pair at baseline
generation, $m_{\mu\mu,p}(\alpha)$ is its invariant mass after the steering is done, and $\boldsymbol{\xi}_{\mathcal{C}}$ is
the steering vector of class $\mathcal{C}$.
In App.~\ref{app:per_pair} we analyze the prediction from Eq.~\eqref{eq:pair_steering}. 
In the rest of this section, we shall focus on the average one.

Averaging over the pairs of class $\mathcal{C}$, the pole mass
$M$ of the lineshape fit should move as
\begin{equation}
\begin{aligned}
    M(\alpha) &\simeq M(0)\,\exp\left(\alpha \sum_k \bar{\mathbf{G}}_{\mathcal{C}}(t_k)^\top\boldsymbol{\xi}_{\mathcal{C}}(t_k) \right) \\
     &=M(0)\,\exp\left(\alpha \sum_k S_{\mathcal{C}}(t_k) \right)\,, 
    \label{eq:prediction_of_steering}
\end{aligned}
\end{equation}
where $\bar{\mathbf{G}}_{\mathcal{C}}(t)$
is the average gradient of class $\mathcal{C}$ at time $t$,  $S_{\mathcal{C}}(t) = \bar{\mathbf{G}}_{\mathcal{C}}(t)^\top\boldsymbol{\xi}_{\mathcal{C}}(t)$ is the gradient response.
Eq.~\eqref{eq:prediction_of_steering} holds to first order in $\alpha$ if every pair of the class responds by the same amount, since $M$ is a fit parameter and not the class mean.
App.~\ref{app:per_pair} checks how far the individual pairs depart from this.
Fig.~\ref{fig:ladder}a compares it with the fit.
We note that how much the invariant mass moves depends on the size of the gradients, the steering vector, and their
cosine similarity, which is defined as
\begin{equation}
    \cos\left(\bar{\mathbf{G}}_{\mathcal{C}}(t_k), \boldsymbol{\xi}_{\mathcal{C}}(t_k)\right)
    = \frac{\bar{\mathbf{G}}_{\mathcal{C}}(t_k)^\top\boldsymbol{\xi}_{\mathcal{C}}(t_k)}
           {\lVert \bar{\mathbf{G}}_{\mathcal{C}}(t_k) \rVert\,\lVert \boldsymbol{\xi}_{\mathcal{C}}(t_k) \rVert}\,,
\label{eq:cosine}
\end{equation}
where, if the $\cos$ were to be 0, it means the gradients and the steering vector are perpendicular, 1 means they are
parallel, and $-1$ that they are anti-parallel.

We perform a fit of the prediction of Eq.~\eqref{eq:prediction_of_steering}, $M(\alpha) = M(0)\,\exp(\alpha S)$, where $S$ is a free parameter. 
$M(0)$ and $ M(\alpha)$ are the values obtained from the lineshape fits at baseline and steered, respectively, for $-3 \leq \alpha \leq 3$.
We find $S_{\mathrm{fit}} = 0.1255$, close to the value of $\sum_k S_Z(t_k) = 0.1220$ computed from the gradients. The fit gives an $R^2 = 0.9998$.

We plot the prediction from Eq.~\eqref{eq:prediction_of_steering}, and the result from the fitted pole mass in Fig.~\ref{fig:ladder}a, where they stay within $5\%$ for all values of $\alpha$ we scanned.
For the width, $\Gamma$, we make an ansatz. Let us take two pairs at the border of the resonance, $m_{\mu\mu}(0) = M(0) \pm \Gamma(0)/2$.
Steering both points (Eq.~\eqref{eq:prediction_of_steering}), and subtracting them, should give us how the steering affects the width,
\begin{equation}
    \Gamma(\alpha) \simeq \Gamma(0)\,\exp\left(\alpha \sum_k S_{\mathcal{C}}(t_k)\right)\,.
\label{eq:width-prediction}
\end{equation}
We plot this ansatz and the result from the fit in Fig.~\ref{fig:ladder}b, where we see completely different results.
Of course, this prediction is related to the full spread seen in Fig.~\ref{fig:spectrum}, rather than the physical width of the $Z$ boson itself.
We already find in App.~\ref{app:fitdetails} that the fitted detector resolution and the fitted width are strongly correlated.
We attribute the extra broadening to muon pairs responding differently to the steering (see App.~\ref{app:per_pair}).
If every pair around the peak responded by the same amount, Eq.~\eqref{eq:width-prediction} would hold to first order in $\alpha$.

The final quantity to consider when assessing the quality of a steering vector is the largest FKT eigenvalue, $\lambda_{\mathcal{C}}(t_k)$, of Eq.~\eqref{eq:eigen}. As we stated before in Sec.~\ref{sec:method}, the larger the eigenvalue, the more indicative it is that the vector will point only at a specific class.

\begin{table}[t]
\caption{The $Z$ steering vector at each block $L$. $\sum_k S_{Z}(t_k)$ is the first-order shift of
$\log m_{\mu\mu}$ of the $Z$ per unit $\alpha$ (Eq.~\eqref{eq:prediction_of_steering}), and $\langle\cdot\rangle_k$ is the
average over the 200 Euler steps.}
\label{tab:quality_steering}
\centering
\small
\begin{tabular}{cccc}
\toprule
$L$ & $\sum_k S_{Z}(t_k)$ & $\left\langle\cos\left(\bar{\mathbf{G}}_{Z}, \boldsymbol{\xi}_{Z}\right)\right\rangle_k$ & $\langle\lambda_{Z}\rangle_k$ \\
\midrule
0 & $1.4\times10^{-1}$ & $1.4\times10^{-1}$ & 0.92 \\
1 & $1.2\times10^{-1}$ & $1.5\times10^{-1}$ & 0.91 \\
2 & $7.7\times10^{-2}$ & $1.4\times10^{-1}$ & 0.91 \\
3 & $4.2\times10^{-2}$ & $1.1\times10^{-1}$ & 0.93 \\
4 & $1.1\times10^{-2}$ & $5.1\times10^{-2}$ & 0.95 \\
5 & $2.6\times10^{-5}$ & $2.4\times10^{-4}$ & 0.93 \\
\bottomrule
\end{tabular}
\end{table}

For the $Z$ boson, Table~\ref{tab:quality_steering} presents, for all blocks, the average or total values of these quantities over all flow time. Fig.~\ref{fig:quality_steering} shows the quality features as a function of $t$.
We note how $\sum_k S_{Z}(t_k)$ is the largest on earlier blocks, meaning that the steering vector will move the $Z$ peak more.
At $L = 0$ it is the largest, and therefore it should move the $Z$ boson the most; however, we found that steering that block also moved the rest of the invariant mass spectrum in ways we did not anticipate, such as distortions of the $J/\psi$ and $\Upsilon$ peaks.
We tested steering with random vectors on $L = 0$ and found that a random steering vector presented modifications to the whole generation output, meaning that block 0 is particularly sensitive to any steering (details in App.~\ref{app:otherblocks}).\footnote{We hypothesize that the gradients are larger on shallower blocks because each block normalises its input, while the norm of the residual stream grows with depth~\citep{shleifer-2021-normformer, sun-2025-curse}.}

$L = 1$ did not present such difficulties, and, furthermore, it is the block with the largest average cosine similarity, and the second highest gradient response.
The average eigenvalue, $\langle\lambda_{Z}\rangle_k$, is the lowest in $L = 1, 2$ (see Fig.~\ref{fig:quality_steering}c for how it changes with $t$).
However, all blocks have values $\langle\lambda_{Z}\rangle_k > 0.9$, meaning that the FKT was quite efficient at distinguishing the $Z$ direction from the rest of the classes.
For reference, we also present the same analysis for $L = 3$ in App.~\ref{app:otherblocks}.

\subsection{Ablation of the steering vector}
We now test the role that $\boldsymbol{\xi}_Z(t)$ has inside of the network. 
To do this causally, we perform an ablation of the output of block $L = 1$ of the muon tokens. 
The ablation has the shape
\begin{align}
    \mathbf{h}_1^{\mu^\pm}(\mathbf{x}_t, t) \to{}& \mathbf{h}_1^{\mu^\pm}(\mathbf{x}_t, t)
    \label{eq:ablation}
    \\
    &- \left[ \boldsymbol{\xi}_Z(t)^\top \mathbf{h}_1^{\mu}(\mathbf{x}_t, t)
    - \langle \boldsymbol{\xi}_Z^\top \mathbf{h}_1^{\mu} \rangle(t) \right] \boldsymbol{\xi}_Z^{\pm}(t)\,,
    \notag
\end{align}
where $\mathbf{h}_1^{\mu} = \left[\mathbf{h}_1^{\mu^-}, \mathbf{h}_1^{\mu^+}\right]$ and
$\boldsymbol{\xi}_Z = \left[\boldsymbol{\xi}_Z^{-}, \boldsymbol{\xi}_Z^{+}\right]$ stack the two muon tokens, and we define
\begin{equation}
    \langle \boldsymbol{\xi}_Z^\top \mathbf{h}_1^{\mu} \rangle(t) = \frac{1}{\mathcal{N}} \sum_{p=1}^{\mathcal{N}}
    \boldsymbol{\xi}_Z(t)^\top \mathbf{h}_{1,p}^{\mu}(\mathbf{x}_t, t)\,,
    \label{eq:ablation-mean}
\end{equation}
the mean over all $\mathcal{N}$ muon pairs of the unsteered generation, where we used all the pairs of all classes in Table~\ref{tab:classes}.

The first term inside of the square brackets of Eq.~\eqref{eq:ablation} subtracts the component of the residual stream along $\boldsymbol{\xi}_Z$, and the second term adds back the average over all pairs, to ensure that there is as little steering as possible~\citep{belrose-2023-leace, li-2024-optimal}. 

\begin{figure}
    \centering
    \includegraphics[width = \linewidth]{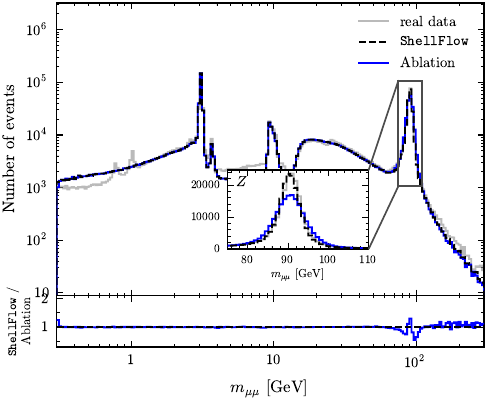}
    \caption{Result of the generated muon spectrum from the ablation in Eq.~\eqref{eq:ablation} (blue), compared with the output from \texttt{ShellFlow} (black, dashed), and real data (grey). 
    The inset shows the $Z$ peak in linear scale, and the lower panel the ratio of \texttt{ShellFlow} to the ablation.}
    \label{fig:ablation}
\end{figure}

We examined the use of this component of the residual stream by analyzing the generated output. 
Ideally, if $\boldsymbol{\xi}_Z$ is a component truly used by the network, it should alter the $Z$ peak only. 
We find that the generated output results in a $\sim 18$\% broader peak.
We show the result of the ablation in Fig.~\ref{fig:ablation}. 
$\mathcal{R}_{60} = 0.012$ between the ablated sample and \texttt{ShellFlow}, below the noise sampling error of 0.018 to 0.034.
The same ablation along random directions broadens the $Z$ peak by only 1--4\%.

Repeating the procedure on a continuum window without a resonance, of the same log-width as the $Z$ window (29.56--36.84~GeV), gives a steering about half as strong, and ablating that direction leaves the spectrum unchanged.

\section{Discussion and conclusions}
\label{sec:limitations}
We have only tested this method for one particle type, muons, where we only changed one observable, the pole mass.
Moreover, since we are working with real data, there are no truth-level classes. We only defined the classes by doing simple cuts in the invariant mass spectrum, 
and not a proper background analysis. Such background analysis is also missing from the $Z$ lineshape fit. 

Despite these limitations, we would like to quantify this gradient-based way of obtaining steering vectors. 
The method we described can also be used in different generative models, not only flow matching models, but also diffusion models, normalizing flows, or Schrödinger bridges; as well as one-shot generative models like generative adversarial networks (GANs) or variational autoencoders (VAEs).

We would also like to highlight the benefit of using steering vectors in physics-based models, as well as any model whose output is quantitative. The fact that we can write the fit of the lineshape of the $Z$ boson allowed us to check how the steering affected not only its mass, but also other physical parameters, such as its width or the detector resolution. 
This is something that image, speech, or language generative models lack, a baseline that can be used to quantify the effects of steering vectors.

For particle physics, a steering vector is a handle on one physical quantity inside a generative model trained on real data.
It tests what the model learned: when we move the $Z$ pole, the transverse momentum of the muons follows the pole at $M/2$, as the decay kinematics require, while the $J/\psi$ and $\Upsilon$ peaks stay in place.
It could also replace some of the samples that analyses generate to build templates and estimate systematic uncertainties, such as variations of a resonance mass, the momentum scale, or the detector resolution, which today need a new simulation or a retrained model.
Our steering vector is not yet such a tool, because the $Z$ peak also broadens as the pole moves; directions that move the pole, the width, and the resolution separately would be.

\section*{Acknowledgements}
We acknowledge the work of the ATLAS Collaboration to record or simulate, reconstruct, and distribute the Open Data
used in this paper, and to develop and support the software with which it was analysed. 
For the beginning and for most of the project, the GPU compute (NVIDIA A100) was provided by the NVIDIA Academic Grant Program.
The last steps of the project used NVIDIA B200 GPUs from the UCloud interactive HPC system, which is managed by the eScience Center at the University of Southern Denmark.
This work was supported by a research grant
(VIL57416) from VILLUM FONDEN.

\section*{AI usage}
Large language models were used in the preparation of this manuscript to revise prose, correct grammar, and write portions of the text. 
Most of the code and visualization was generated with the assistance of large language models, primarily through the Claude Code agentic coding environment (Anthropic).
All output was reviewed and validated by the authors.

\section*{Data and code availability}
The code that reproduces the results shall be available at \url{https://github.com/kurquia/shellsteer}.
\texttt{ShellFlow}, \texttt{ShellFlow-B}, and \texttt{ShellGATr} are available at \url{https://huggingface.co/pomidori73/ShellFlow}.
The data are ATLAS Open Data released by the ATLAS Collaboration~\citep{atlas-2025-opendata}, available at \url{https://opendata.cern.ch/record/93910}.

\bibliography{refs}
\bibliographystyle{icml2026_links}

\newpage
\appendix
\onecolumn

\section{Derivation of the linear response}
\label{app:derivation}
Let's take again Eq.~\eqref{eq:log_mass} at the end of generation
\begin{equation}
    \log m_{\mu\mu} = \log E_{\mathrm{ref}} + \frac{1}{2}\left(y_{E_{+}} + y_{E_{-}}\right) + \frac{1}{2}\log\left[2\,(1 - \hat{\mathbf{n}}_{+}\cdot\hat{\mathbf{n}}_{-})\right]\,,
\end{equation}
its dependence on the block outputs becomes apparent once we consider how $y_{E}$ and $\hat{\mathbf{n}}$ depend on the velocity field, which are both dependent on the outputs of the blocks themselves
\begin{align}
    y_{t + \Delta t} &= y_t + \Delta t\,\dot{y}_t^{\theta}(\mathbf{x}_t)\,, \\
    \hat{\mathbf{n}}_{t + \Delta t} &= \exp_{\hat{\mathbf{n}}_t}\left(\Delta t\,\dot{\hat{\mathbf{n}}}_t^{\theta}(\mathbf{x}_t)\right)\,,
\end{align}
where $\mathbf{x}_t$ is the state of all particles of the event, and the exponential map for $S^2$ is
\begin{equation}
    \exp_{\mathbf{x}}\left(\mathbf{u}\right) = 
\mathbf{x}\cos(\left\|\mathbf{u} \right\|) + \frac{\mathbf{u}}{\left\| \mathbf{u}\right\|} \sin(\left\| \mathbf{u}\right\|)\,.
\end{equation}

Then, $\log m_{\mu\mu}$ is a function of the velocity field for all Euler steps. 
At each step, the velocity field is itself the output of all blocks. 
Adding $\boldsymbol{\delta}^{\pm}(t_k)$ to the output of block $L$ on the muon tokens gives
\begin{equation}
    \log m_{\mu\mu} = f\left(\mathbf{h}_\ell\left(\mathbf{h}_{\ell-1}\left(\cdots \mathbf{h}_{L+1}\left(\mathbf{h}_L^{\mu^\pm}\left(\mathbf{h}_{L-1}\left(\cdots \mathbf{h}_0\right)\right) + \boldsymbol{\delta}^{\pm}(t_k)\right)\right)\right)\right)\,,
\end{equation}
where $\mathbf{h}_\ell$ is the last block, and $\boldsymbol{\delta}^{\pm}(t_k)$ is zero on all other tokens. 
We evaluate the network at every Euler step, meaning that $\log m_{\mu\mu}$ is a function of all the $\boldsymbol{\delta}^\pm(t_k)$: $f(\vec{\boldsymbol{\delta}}^{+}, \vec{\boldsymbol{\delta}}^{-})$, with $\vec{\boldsymbol{\delta}}^{\pm} = \left(\boldsymbol{\delta}^{\pm}(t_0), \dots,
\boldsymbol{\delta}^{\pm}(t_{N-1})\right)$, and $f(0, 0)$ is the generation without the edit.
A linear expansion from the usual multivariate Taylor expansion gives
\begin{equation}
    f(\vec{\boldsymbol{\delta}}^{+}, \vec{\boldsymbol{\delta}}^{-}) = f(0, 0) + \sum_{k=0}^{N-1} \left[
    \left.\frac{\mathrm{D} f}{\mathrm{D} \boldsymbol{\delta}^{-}(t_k)}\right|_{\vec{\boldsymbol{\delta}}^{\pm} = 0}^\top \boldsymbol{\delta}^{-}(t_k)
    + \left.\frac{\mathrm{D} f}{\mathrm{D} \boldsymbol{\delta}^{+}(t_k)}\right|_{\vec{\boldsymbol{\delta}}^{\pm} = 0}^\top \boldsymbol{\delta}^{+}(t_k)
    \right]
    + \mathcal{O}\left(\left\| \vec{\boldsymbol{\delta}}^\pm\right\|^2\right)\,.
\end{equation}
This total derivative, of course, is just the derivative with respect to the block outputs, what in the main text is
referred to simply as \textit{gradients}
\begin{equation}
    \frac{\mathrm{D} f}{\mathrm{D} \mathbf{h}_L^{\mu^{\pm}}(t_k)} = \left.\frac{\mathrm{D} f}{\mathrm{D} \boldsymbol{\delta}^\pm(t_k)}\right|_{\vec{\boldsymbol{\delta}}^\pm = 0}\,.
\end{equation}

Now, setting $\boldsymbol{\delta}^{\pm}(t_k) = \alpha\,\boldsymbol{\xi}^{\pm}(t_k)$, and writing explicitly the time dependence we arrive at
\begin{equation}
\log m_{\mu\mu}(\alpha) = \log m_{\mu\mu}(0) + \alpha\,\sum_{k = 0}^{N-1} \left[\frac{\mathrm{D} \log m_{\mu\mu}(0)}{\mathrm{D} \mathbf{h}_L^{\mu^{-}}(t_k)} ^\top \boldsymbol{\xi}^-(t_k) + \frac{\mathrm{D} \log m_{\mu\mu}(0)}{\mathrm{D} \mathbf{h}_L^{\mu^{+}}(t_k)} ^\top \boldsymbol{\xi}^+(t_k) \right] + \mathcal{O}\left(\alpha^2\right)\,,
\end{equation}
which is exactly what we have in Eq.~\eqref{eq:linresp}.

\section{Steering vectors from different methods}
\label{app:different_methods}
\begin{figure}[t]
\centering
\includegraphics[width = \textwidth]{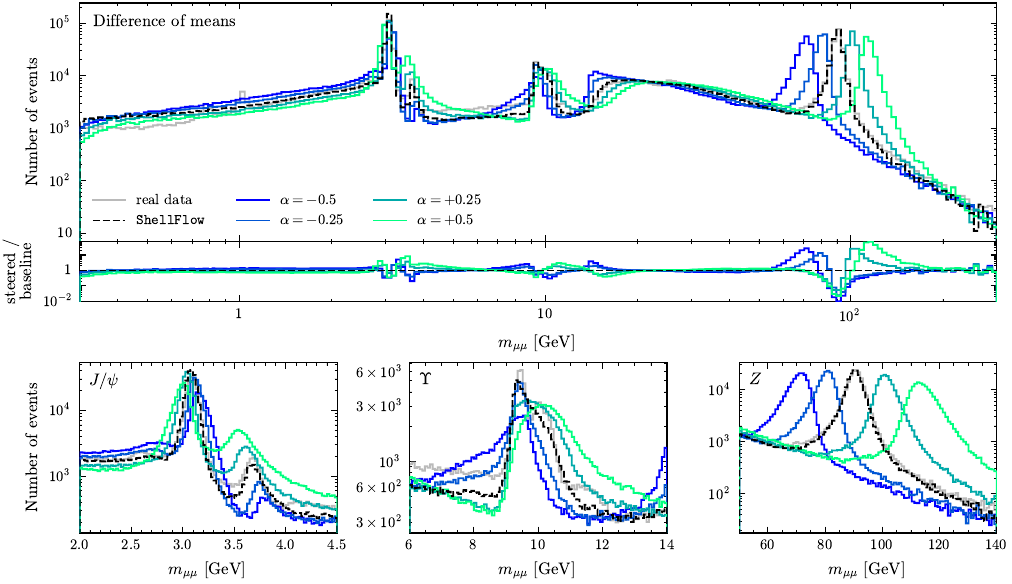}
\caption{The steering of the $Z$ boson with the difference-of-means steering vector, as in Fig.~\ref{fig:spectrum}.
Top: dimuon invariant mass of one million muon pairs, compared with real data (grey), baseline generation of
\texttt{ShellFlow} (black, dashed), and the results from adding the steering vector at block 1 with $\alpha = \pm 0.25, \pm 0.5$.
Middle: steered over baseline. Bottom: zoom-in over the $J/\psi$, $\Upsilon$, and $Z$ peaks.}
\label{fig:app_spectrum_meandiff}
\end{figure}

\begin{figure}[t]
\centering
\includegraphics[width = \textwidth]{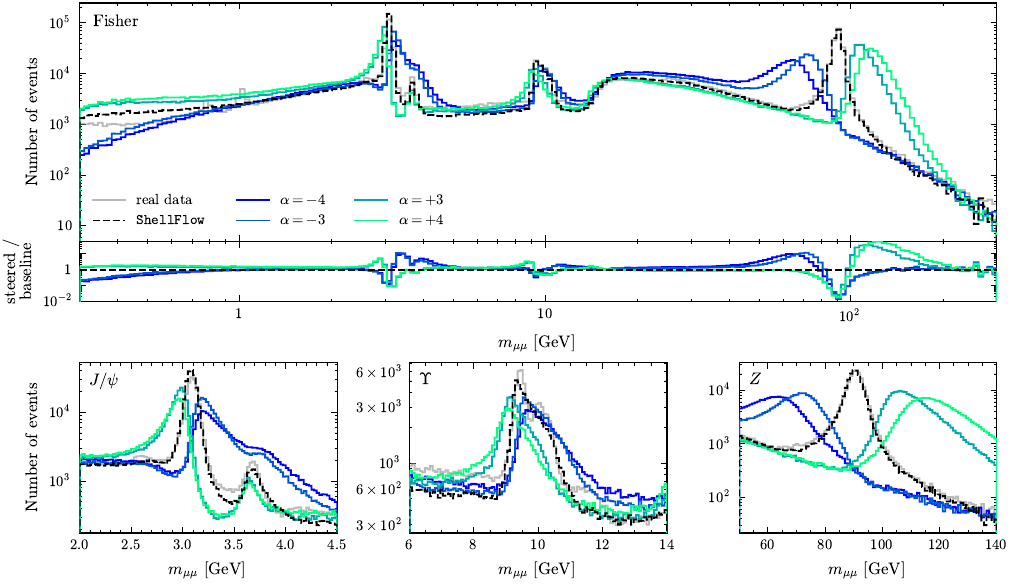}
\caption{The same as Fig.~\ref{fig:app_spectrum_meandiff} for the Fisher steering vector, with
$\alpha = \pm 3, \pm 4$ (blue to green).}
\label{fig:app_spectrum_fisher}
\end{figure}
We consider two different criterion functions to the one defined in Eq.~\eqref{eq:fk-criterion}.
We use them as baselines for the steering vectors of the main text.
All three criteria act on the same unit gradients, $\hat{\mathbf{G}}_p(t_k)$, at every step $t_k$ (we omit the dependence
on $t_k$), with the same classes $\mathcal{C}$ and $\mathcal{C}^\prime$ as in Sec.~\ref{sec:directions}.

The criteria are written in terms of the means, second moments, and covariances of the classes $\mathcal{C}$ and
$\mathcal{C}^\prime$,
\begin{equation}
\mathbf{m}_{\mathcal{C}} = \frac{1}{n_{\mathcal{C}}}\sum_{p\in\mathcal{C}} \hat{\mathbf{G}}_p\,,\qquad
\boldsymbol{\Sigma}_{\mathcal{C}} = \frac{1}{n_{\mathcal{C}}}\sum_{p\in\mathcal{C}} \hat{\mathbf{G}}_p\,\hat{\mathbf{G}}_p^{\top}\,,\qquad
\mathbf{K}_{\mathcal{C}} = \boldsymbol{\Sigma}_{\mathcal{C}} - \mathbf{m}_{\mathcal{C}}\,\mathbf{m}_{\mathcal{C}}^{\top}\,,
\label{eq:class-moments}
\end{equation}
and similar for $\mathcal{C}^\prime$, where every class in $\mathcal{C}^\prime$ enters with the same weight, as in
Eq.~\eqref{eq:moments}.\footnote{Here $\boldsymbol{\Sigma}_{\mathcal{C}^\prime}$ is the average over the three classes in
$\mathcal{C}^\prime$, rather than the sum of Eq.~\eqref{eq:moments}. The factor of three does not change the
maximum of Eq.~\eqref{eq:fk-criterion}.}

The three criterion functions are\footnote{We also experimented with a criterion function, $J = \boldsymbol{\xi}^\top\mathbf{m}_\mathcal{C}\mathbf{m}_\mathcal{C}^\top\boldsymbol{\xi} / \boldsymbol{\xi}^\top\boldsymbol{\Sigma}_{\mathcal{C}^\prime}\boldsymbol{\xi}$, to minimize the spread of the $Z$ peak observed in the main text. However, this did not give results better than the FK criterion, with similar numbers in $\mathcal{R}_{60}$ and broadening.}
\begin{align}
J_{D} &= \frac{\boldsymbol{\xi}^{\top}\left(\mathbf{m}_{\mathcal{C}} - \mathbf{m}_{\mathcal{C}^\prime}\right)
               \left(\mathbf{m}_{\mathcal{C}} - \mathbf{m}_{\mathcal{C}^\prime}\right)^{\top}\boldsymbol{\xi}}
              {\boldsymbol{\xi}^{\top}\boldsymbol{\xi}}\,,
      && \text{difference of means}\,, \label{eq:J-D}\\[1ex]
J_{F} &= \frac{\boldsymbol{\xi}^{\top}\left(\mathbf{m}_{\mathcal{C}} - \mathbf{m}_{\mathcal{C}^\prime}\right)
               \left(\mathbf{m}_{\mathcal{C}} - \mathbf{m}_{\mathcal{C}^\prime}\right)^{\top}\boldsymbol{\xi}}
              {\boldsymbol{\xi}^{\top}\left(\mathbf{K}_{\mathcal{C}} + \mathbf{K}_{\mathcal{C}^\prime}\right)\boldsymbol{\xi}}\,,
      && \text{Fisher criterion}\,, \label{eq:J-F}\\[1ex]
J_{FK} &= \frac{\boldsymbol{\xi}^{\top}\boldsymbol{\Sigma}_{\mathcal{C}}\,\boldsymbol{\xi}}
               {\boldsymbol{\xi}^{\top}\left(\boldsymbol{\Sigma}_{\mathcal{C}} + \boldsymbol{\Sigma}_{\mathcal{C}^\prime}\right)\boldsymbol{\xi}}\,,
      && \text{Fukunaga--Koontz criterion}\,. \label{eq:J-FK}
\end{align}
The difference of means looks for the direction along which the means of the two classes are farthest apart. The Fisher
criterion~\citep{fisher-1936-discriminant,fisher-1938-statistical,rao-1948-utilization} looks for the direction that
separates the means while keeping the spread of each class small. The Fukunaga--Koontz criterion of
Eq.~\eqref{eq:fk-criterion} looks for the direction where the second moment of $\mathcal{C}$ is large, while that of
$\mathcal{C}^\prime$ is small.

All three are generalized Rayleigh quotients,
$J = \frac{\boldsymbol{\xi}^{\top}\mathbf{A}\,\boldsymbol{\xi}}{\boldsymbol{\xi}^{\top}\mathbf{B}\,\boldsymbol{\xi}}$, with
$\mathbf{A}$ and $\mathbf{B}$ symmetric and $\mathbf{B}$ positive definite. The maximum solves the generalized eigenproblem
\begin{equation}
\mathbf{A}\,\boldsymbol{\xi} = \lambda\,\mathbf{B}\,\boldsymbol{\xi}\,,
\label{eq:gen-eigen-AB}
\end{equation}
which becomes an ordinary eigenproblem after a whitening transformation of $\mathbf{B}$, as we did for the
Fukunaga--Koontz criterion in Sec.~\ref{sec:directions}.
For $J_D$ and $J_F$, $\mathbf{A}$ has rank one, so Eq.~\eqref{eq:gen-eigen-AB} has a single nonzero eigenvalue,
and its unit eigenvector has the closed form
\begin{equation}
\boldsymbol{\xi}_{D} = \frac{\mathbf{m}_{\mathcal{C}} - \mathbf{m}_{\mathcal{C}^\prime}}
                            {\left\lVert\mathbf{m}_{\mathcal{C}} - \mathbf{m}_{\mathcal{C}^\prime}\right\rVert}\,,\qquad
\boldsymbol{\xi}_{F} = \frac{\left(\mathbf{K}_{\mathcal{C}} + \mathbf{K}_{\mathcal{C}^\prime}\right)^{-1}
                             \left(\mathbf{m}_{\mathcal{C}} - \mathbf{m}_{\mathcal{C}^\prime}\right)}
                            {\left\lVert\left(\mathbf{K}_{\mathcal{C}} + \mathbf{K}_{\mathcal{C}^\prime}\right)^{-1}
                             \left(\mathbf{m}_{\mathcal{C}} - \mathbf{m}_{\mathcal{C}^\prime}\right)\right\rVert}\,.
\label{eq:xi-D-F}
\end{equation}
We fix their sign as for the Fukunaga--Koontz direction.
The difference of means is the most common way to build a steering vector in language
models~\citep{li-2023-iti,rimsky-2024-caa,arditi-2024-refusal}, and it is a strong baseline among steering
methods~\citep{wu-2025-axbench}.

\begin{wrapfigure}{r}{0.455\textwidth}
\centering
\includegraphics[width = 0.45\textwidth]{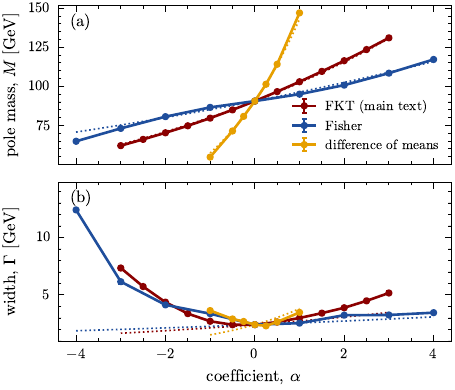}
\caption{Fitted pole (a) and width (b) of the $Z$ boson against $\alpha$ for the three steering vectors at block 1,
as in Fig.~\ref{fig:ladder}; dotted: the first-order prediction of each. The Fisher steering vector skews the $Z$ peak,
so its fits describe the peak less well.}
\label{fig:app_ladder_methods}
\vspace{-12pt}
\end{wrapfigure}
Neither of the two is optimal for a localized shift,
where we are maximizing the response of the pairs in $\mathcal{C}$ while minimizing that of the pairs in $\mathcal{C}^\prime$.
The difference of means only optimizes the mean responses, so the responses of single pairs can be large and cancel in the mean.
The Fisher criterion keeps the spread of $\mathcal{C}^\prime$ small, but not its mean response, so $\mathcal{C}^\prime$ can shift as a whole.
Only $J_{FK}$ minimizes both the spread and the mean response of $\mathcal{C}^\prime$.

\subsection{Steering of the \texorpdfstring{$Z$}{Z} boson}
\label{app:methods_steering}
We steer the $Z$ boson with the steering vectors of Eq.~\eqref{eq:xi-D-F} exactly as in
Sec.~\ref{sec:experiments}.
We chose the values of $\alpha$ so that the steering moves the $Z$ as much as the FK steering vector does at $\alpha = \pm 1, \pm 2$.
For difference of means, this is $\alpha = \pm 0.25, \pm 0.5$, and for Fisher this is $\alpha = \pm 3, \pm 4$.

Neither method is selective.
We can see in Fig.~\ref{fig:app_spectrum_meandiff} and Fig.~\ref{fig:app_spectrum_fisher} that the steering succeeds in shifting the $Z$ pole, while at the same time altering the rest of the continuum and the other resonances. 
This is in contrast with what we have in the main text. 
$\mathcal{R}_{60} = 0.112$ for the values of $\alpha$ considered for FK. In contrast for difference of means it is 0.363 and for Fisher 0.718.
All these values are obtained for $\alpha = -2, -0.5, -4$ respectively for FK, difference of means, and Fisher.

As for the predictions given in Eq.~\eqref{eq:prediction_of_steering}, we perform the same fit as in Sec.~\ref{sec:steering_shape} and find that difference of means has an $R^2 = 0.9992$ and for Fisher $R^2 = 0.979$, both following the exponential prediction. 
The fitted values of $\sum S_Z(t_k)$ are close to the ones measured from the gradients.
For difference of means the value of the sum (fitted) is 0.4569 (0.4871), and for Fisher 0.0617 (0.0682).

\section{Steering of other networks}
\label{app:othernetworks}

\begin{table}[t]
\centering
\caption{The three networks. \texttt{ShellGATr} counts its channels in four-vectors (v) and scalars (s).
FLOPs: one forward pass per particle, measured at 8 particles per event.}
\label{tab:networks}
\footnotesize
\setlength{\tabcolsep}{3.3pt}
\begin{tabular}{lccccccccccc}
\toprule
 & \multicolumn{2}{c}{Channels} & \multicolumn{4}{c}{Architecture} & \multicolumn{5}{c}{Training} \\
\cmidrule(lr){2-3}\cmidrule(lr){4-7}\cmidrule(lr){8-12}
 & embedding & MLP & heads & blocks & parameters & FLOPs & data & events & epochs & lr & GPUs \\
\midrule
\texttt{ShellFlow}   & 128 & 512 & 4 & 6 & 3.02M & 5.5M & \begin{tabular}[c]{@{}c@{}}\texttt{1LMET30} +\\[-0.25ex] \texttt{2to4lep}\end{tabular} & 783M & 30 & $6 \cdot 10^{-4}$ & 8 A100 \\[2ex]
\texttt{ShellFlow-B} & 128 & 512 & 4 & 6 & 3.02M & 5.5M & \texttt{2to4lep} & 268M & 100 & $3 \cdot 10^{-4}$ & 4 A100 \\[2ex]
\texttt{ShellGATr}   & 32\,v + 96\,s & 128\,v + 384\,s & 4 & 6 & 3.09M & 6.2M & \texttt{2to4lep} & 268M & 100 & $3 \cdot 10^{-4}$ & 4 B200 \\
\bottomrule
\end{tabular}
\end{table}

We test our method for finding steerable features on two different versions of \texttt{ShellFlow}. 
Both have around the same number of trainable parameters, but were trained only on the \texttt{2to4lep} selection with 268M events.
For comparison, the model that is the protagonist of the main text was trained on \texttt{1LMET30 + 2to4lep}, that merges two ATLAS selections, and has 783M events.

We introduce \texttt{ShellFlow-B}, an early iteration of \texttt{ShellFlow} that was trained during the preparation of \citet{kato-2026-shellflow}.
We also present \texttt{ShellGATr}, which uses the L-GATr-slim layers, a Lorentz-equivariant transformer, rather than the \textit{classical} transformer \citep[see][]{spinner-2024-lgatr, petitjean-2025-economical}.
\texttt{ShellGATr} was trained specifically to test steering, but we found interesting features, such as the fact that it could separate the $\Upsilon(1\mathrm{S})$ from the $\Upsilon(2\mathrm{S})$, which the other networks could not do. 
\texttt{ShellGATr} will be the focus of a future study.

Table~\ref{tab:networks} summarizes the differences in the network architectures and during training. 
Since \texttt{ShellFlow} is trained on a different dataset, its manifold coordinates are normalized differently from those of the other two networks.
Furthermore, in \texttt{ShellFlow-B} and \texttt{ShellGATr}, each particle type has its own input and output layers.
Both \texttt{ShellFlow} and \texttt{ShellFlow-B} directly read the manifold coordinates, whereas \texttt{ShellGATr} has a layer that converts those coordinates into four-vectors
\begin{equation}
    \label{eq:four_vector_shellgatr}
    \begin{aligned}
    \text{Massless embedding:}&\qquad& 
    p^\nu &= E_\mathrm{ref}\,e^{y_E}\,\left(1, \hat{\mathbf{n}}\right)\,, \\
    \text{Massive embedding:}&\qquad& 
    p^\nu &= m\,\left(\cosh\chi, \sinh\chi\,   \hat{\mathbf{n}}\right)\,,
\end{aligned}
\end{equation}
where $\chi$ is the boost rapidity and $m$ is the mass of the particle.
Each \texttt{ShellGATr} token carries $p^\nu$ together with two fixed four-vectors, the time axis $\hat{t} = (1, 0, 0, 0)$ and the beam axis $\hat{z} = (0, 0, 0, 1)$, which a learned linear map mixes into the vector channels.
The chart coordinates $y_E$ or $\chi$, $n_z$ (the beam component), and $\log m$ enter as scalars.

The missing transverse momentum enters the cross-attention as plain features in \texttt{ShellFlow} and \texttt{ShellFlow-B}, and as a four-vector together with its magnitude in \texttt{ShellGATr}.

For \texttt{ShellGATr}, we chose the activation functions used in \citet{petitjean-2025-economical}. 
For vectors it is $\mathrm{sigmoid}\left(\frac{1}{2}\langle g_{1,C},g_{2,C}\rangle\right)\,v_C$ where $g_{i,C}, v_C$ are linear combinations of the input vectors, and $\langle\cdot,\cdot\rangle$ is the product in Minkowski space.
For scalars it is $\mathrm{GELU}(W_1\,s) \odot W_2\,s$, where $s$ are input scalars, and $W_i$ are trainable parameters.
The activation functions of \texttt{ShellFlow-B} are the same as the \texttt{ShellFlow} ones (SwiGLU).

Finally, we have differences that these two new networks have at the level of loss.
The flow-matching loss weights each particle by a quality score built from its detector features \citep[Eq.~16]{kato-2026-shellflow}.
In \texttt{ShellFlow}, this weight has a floor of 0.5, whereas in the other two networks the floor is 0.
The auxiliary loss is weighted by how far the generated distribution is from real data \citep[Eq.~20]{kato-2026-shellflow}.
This weight has a ceiling of 3 in \texttt{ShellFlow} \citep[see Table~8 of][]{kato-2026-shellflow} and of 5 in the other two networks.

\begin{figure}[!h]
\centering
\includegraphics[width = \textwidth]{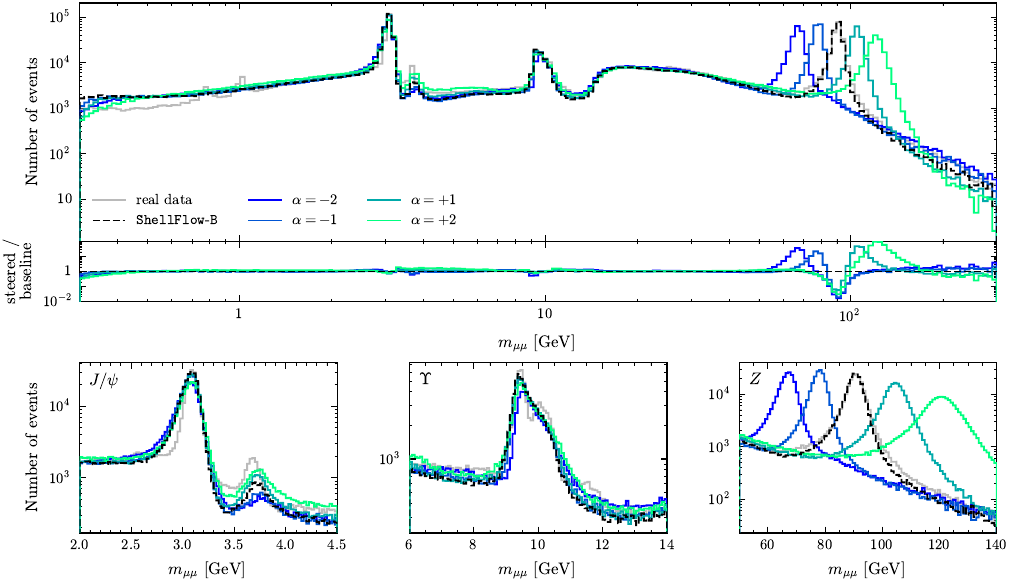}
\caption{Dimuon invariant mass of \texttt{ShellFlow-B} for the $Z$ steering vector at block $L = 1$, for $\alpha = \pm 1, \pm 2$.}
\label{fig:app_spectrum_shellflowb}
\end{figure}

\begin{figure}[!h]
\centering
\includegraphics[width = \textwidth]{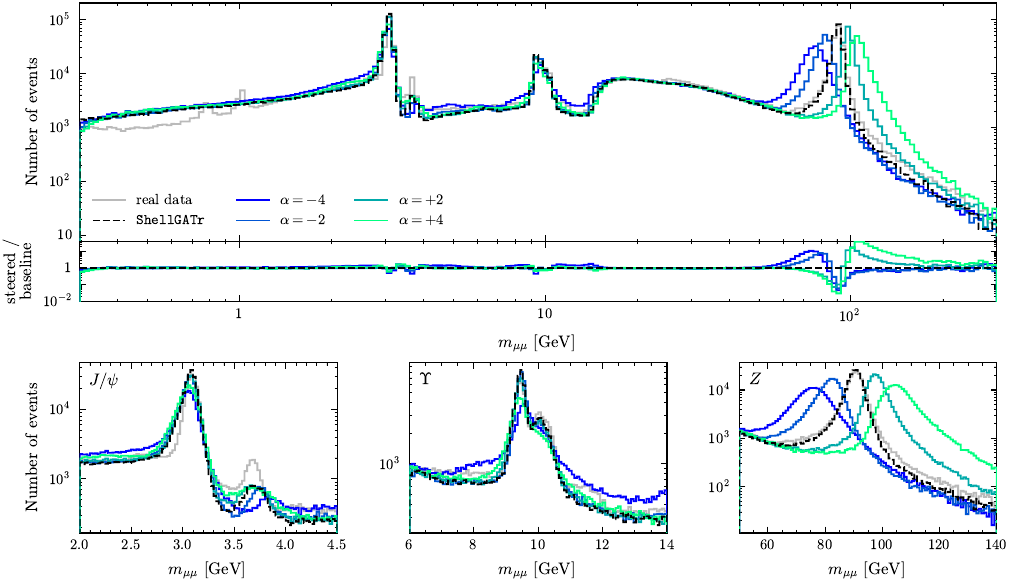}
\caption{The same as Fig.~\ref{fig:app_spectrum_shellflowb} for \texttt{ShellGATr}, for $\alpha = \pm 2, \pm 4$.}
\label{fig:app_spectrum_shellgatr}
\end{figure}

\clearpage
\subsection{Steering of both networks}
\begin{wrapfigure}{r}{0.5\textwidth}
\centering
\includegraphics[width = 0.42\textwidth]{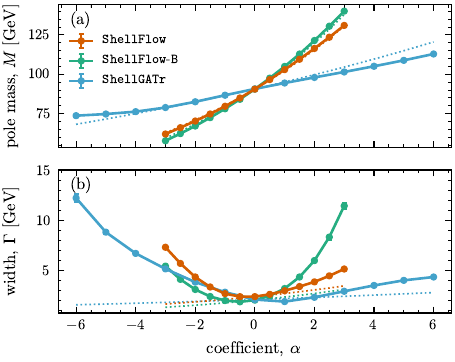}
\caption{Fitted pole (a) and width (b) against $\alpha$ for the $Z$ steering vector at block $L = 1$ of the three networks, as in Fig.~\ref{fig:ladder}, mean of three seeds.
Dotted: the baseline value of each network times $\exp\left(\alpha\sum_k S_{Z}(t_k)\right)$.}
\label{fig:app_ladder_networks}
\end{wrapfigure}
We now perform the same procedure done in the main text. We take two samples of 1,000,000 muon pairs from \texttt{2to4lep} validation events. On the first sample, we compute the gradients and the steering vectors, and on the second we steer.
As in Sec.~\ref{sec:steering_effects}, we use three noise seeds, and within a seed the steered and baseline generations start from the same noise.

We compute the gradients on the $\mu^-$ and $\mu^+$ tokens for block $L = 1$, we concatenate them, and compute their second moments.
The second moments are then separated by classes corresponding to $Z, J/\psi, \Upsilon$, or the continuum, and we do a one-versus-rest FK procedure.

We note that this procedure is not Lorentz-invariant. In \texttt{ShellGATr}, the gradients $\mathbf{G}_p(t_k)$ have both scalar and vector components.
For the vector component, the normalization of the gradients is done with the Euclidean metric, thus breaking Lorentz invariance.
This is not of large concern for us, since the network already breaks the Minkowski group to only have rotations about the beam axis.
Our procedure preserves this $\mathrm{O}(2)$ symmetry.\footnote{We also tried Lorentz-invariant steering vectors, and they altered the rest of the resonances more than the Euclidean ones.}

We show the resulting invariant mass spectra in Figs.~\ref{fig:app_spectrum_shellflowb} and \ref{fig:app_spectrum_shellgatr}.
We find similar results for \texttt{ShellFlow-B} as in the main text, but we found the \texttt{ShellGATr} network to be more sensitive in generation for the rest of the resonances.
For \texttt{ShellFlow-B}, for $m_{\mu\mu} < 60$~GeV, the steered and baseline histograms differ per bin by less than 8\% on average for $|\alpha| \le 1$.
\texttt{ShellGATr} needs a larger $\alpha$ to move the $Z$ by the same amount, and it alters the continuum more.
At $\alpha = -3$, which moves the $Z$ as far as $\alpha = -1$ does in \texttt{ShellFlow-B}, the histograms differ below 60~GeV three times as much as in \texttt{ShellFlow-B}.

\begin{wraptable}{r}{0.5\textwidth}
\caption{The $Z$ steering vector at block $L = 1$ of the three networks, as in Table~\ref{tab:quality_steering}.}
\label{tab:app_quality_networks}
\centering
\footnotesize
\setlength{\tabcolsep}{4pt}
\begin{tabular}{lccc}
\toprule
 & $\sum_k S_{Z}(t_k)$ & $\left\langle\cos\left(\bar{\mathbf{G}}_{Z}, \boldsymbol{\xi}_{Z}\right)\right\rangle_k$ & $\langle\lambda_{Z}\rangle_k$ \\
\midrule
\texttt{ShellFlow}   & 0.122 & 0.147 & 0.91 \\
\texttt{ShellFlow-B} & 0.140 & 0.108 & 0.92 \\
\texttt{ShellGATr}   & 0.047 & 0.111 & 0.91 \\
\bottomrule
\end{tabular}
\end{wraptable}
We fit the $Z$ peak of both networks with the lineshape of App.~\ref{app:lineshapefit}.
Figure~\ref{fig:app_ladder_networks} shows the fitted pole and width against $\alpha$ for the three networks.
For \texttt{ShellFlow-B}, the pole stays within 3\% of the prediction of Eq.~\eqref{eq:prediction_of_steering} for all values of $\alpha$ we scanned, and the fit of $M(\alpha) = M(0)\,\exp(\alpha S)$ gives $S_{\mathrm{fit}}$ close to $\sum_k S_Z(t_k)$.
For \texttt{ShellGATr}, the pole stays within 2\% of the prediction for $-4 \le \alpha \le 2$.
For larger $|\alpha|$ it moves less than predicted, and below $\alpha = -4$ it stays near 75~GeV.

As for \texttt{ShellFlow}, the width of both networks grows faster than the ansatz of Eq.~\eqref{eq:width-prediction}.
\texttt{ShellFlow-B} broadens most for positive $\alpha$, and \texttt{ShellGATr} for negative $\alpha$.
The lineshape fits are also poorer than for \texttt{ShellFlow}: the steered peak of \texttt{ShellGATr} becomes lopsided (Fig.~\ref{fig:app_spectrum_shellgatr}), and its deviance per degree of freedom reaches 41 for positive $\alpha$.

We do not know the cause of these differences.
They do not show in the first-order quantities of Table~\ref{tab:app_quality_networks}.
The steering vector of \texttt{ShellGATr} is as aligned with $\bar{\mathbf{G}}_{Z}$ and as $Z$-specific as those of the other two networks.
Only its $\sum_k S_Z(t_k)$ is smaller, so it needs a larger $\alpha$.
The networks differ in several places that the steering vector passes through,
for example the input of the activation functions of the vector components are \texttt{ShellGATr} are quadratic, whereas the argument of the gates of \texttt{ShellFlow} and \texttt{ShellFlow-B} is linear in it.

\clearpage
\begin{figure*}[!t]
\centering
\includegraphics[width = \textwidth]{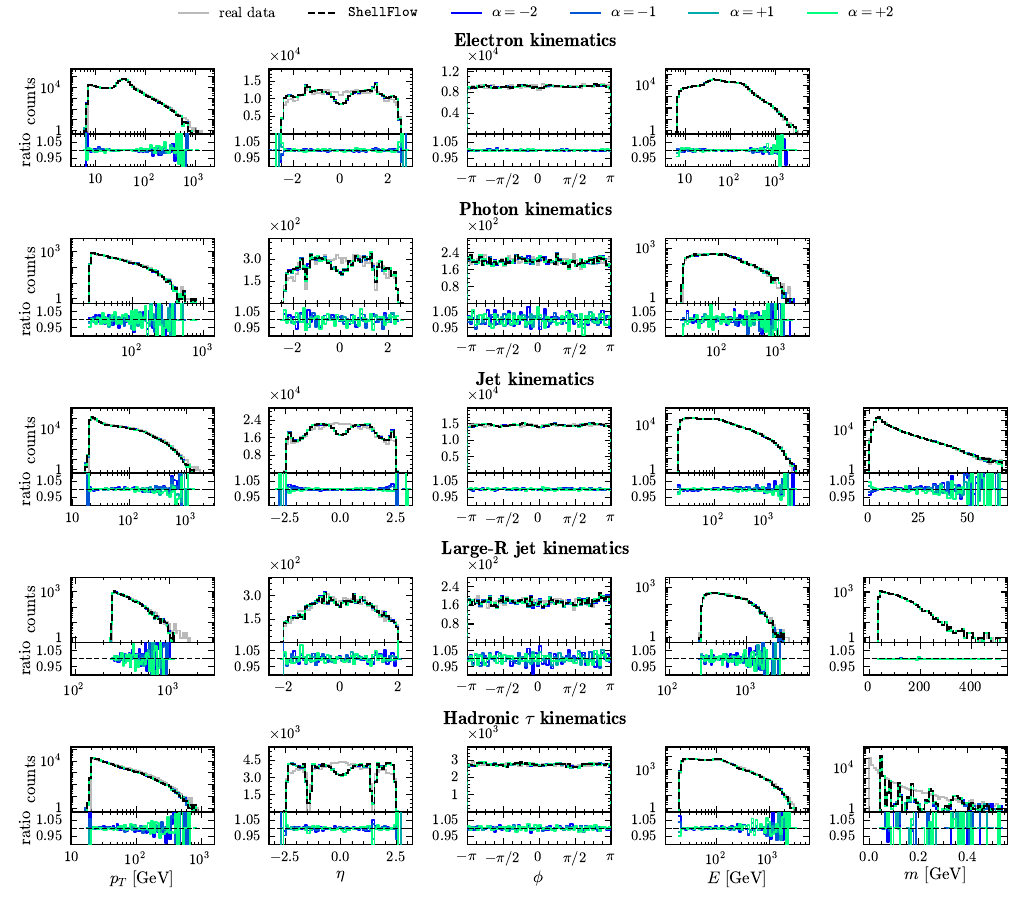}
\caption{Kinematics of the electrons, photons, jets, large-R jets, and hadronic taus
after the steering on muon tokens at block $L = 1$ that moves the $Z$ boson peak.
Same events and steering as in Fig.~\ref{fig:kinematics}: 1,000,000 validation events.
Columns: transverse momentum, pseudorapidity, azimuth, energy, and, for the massive particles, mass; 60 bins per panel.
Real data (grey), baseline generation (black, dashed), and steered with $\alpha = -2, -1, +1, +2$
(blue to green). Ratio panels: steered over baseline.}
\label{fig:app_all_particles}
\end{figure*}

\section{Further results of the steering}
\label{app:steering}

\subsection{Kinematics of the steered events}
\label{app:allparticles}

We now show the results of the steering on reconstructed particles that were not specifically steered; 
that is electrons, photons, hadronic taus, jets, and large-R jets.
Figure~\ref{fig:app_all_particles}
shows the results for every particle with the same steering vector as in
the main text: the $Z$ steering vector on block $L=1$ for $\alpha =  \pm 1, \pm 2$.
We do not expect huge edits to occur for these particles, as, of course, the steering vector is only
added to the muon tokens.
However, the tokens of the rest of the particles should not remain unaffected. 
At every inference step, subsequent attention blocks also read the steering vector, 
thus altering the generated kinematics of all particles in the event.

For $|\alpha| \le 1$, no bin with at least 1,000 baseline particles changes by more than 2.4\%.
The largest changes sit at the edges of the pseudorapidity acceptance and at high transverse momentum, 
where there is low statistics.

\subsection{The \texorpdfstring{$Z$}{Z} steering vector at blocks 0 and 3}
\label{app:otherblocks}

\begin{figure*}[t]
\centering
\includegraphics[width = \textwidth]{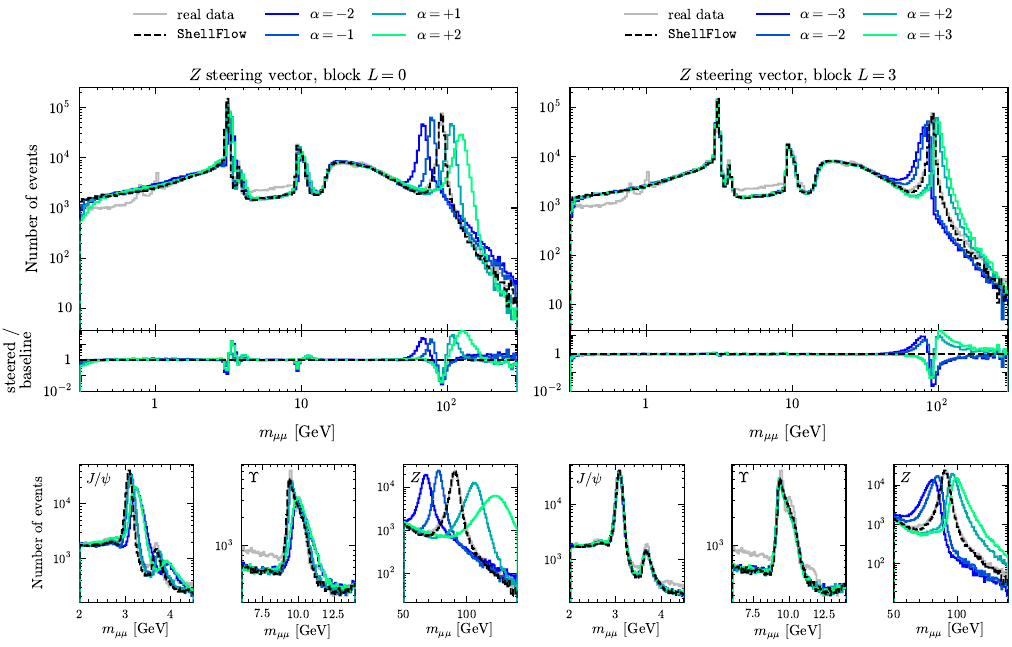}
\caption{The $Z$ steering vector at block $L = 0$ (left, $\alpha = \pm 1, \pm 2$) and $L = 3$ (right,
$\alpha = \pm 2, \pm 3$), as in Fig.~\ref{fig:spectrum}. Top: dimuon invariant mass of one million muon pairs, real
data (grey), baseline generation of \texttt{ShellFlow} (black, dashed), and steered (blue to green). Middle: steered over baseline. Bottom: zooms on the $J/\psi$, $\Upsilon$, and $Z$.}
\label{fig:app_spectrum_blocks}
\end{figure*}

\begin{wrapfigure}[15]{r}{0.4\textwidth}
\centering
\includegraphics[width = 0.45\textwidth]{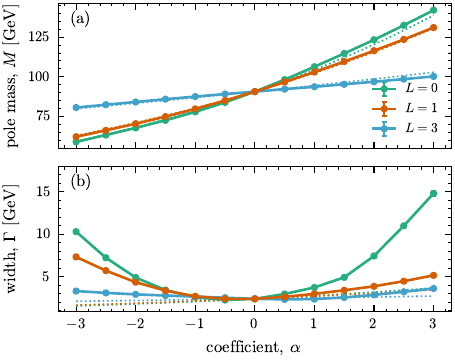}
\caption{Fitted pole and width against $\alpha$ for the $Z$ steering vector at blocks $L = 0$, 1, and 3, as in
Fig.~\ref{fig:ladder}. Dotted: the baseline value times $\exp\left(\alpha\sum_k S_{Z}(t_k)\right)$ of each block
(Table~\ref{tab:quality_steering}).}
\label{fig:app_ladder_blocks}
\end{wrapfigure}

We repeat the same exercises as in the main text for blocks $L = 0$ and $L = 3$. 
Figure~\ref{fig:app_spectrum_blocks} shows the dimuon invariant mass plots for both, and Fig.~\ref{fig:app_ladder_blocks} 
the fitted pole and width against $\alpha$.
The quality features of both blocks are in Table~\ref{tab:quality_steering} and Fig.~\ref{fig:quality_steering}.

As we mentioned in the main text, $L = 0$ is a sensitive block. 
We found that adding the steering vector, as well as a random vector, 
altered the invariant mass spectrum way more than the steering on block $L = 1$. 
This is clear from the peaks in $J/\psi$ and $\Upsilon$ in Fig.~\ref{fig:app_spectrum_blocks}.

At $L = 3$, $\sum_k S_{Z}(t_k)$ is a third of its value at $L = 1$, so the steering vector needs $\alpha = \pm 3$ to
move the $Z$ as far as $L = 1$ does with $\alpha = \pm 1$.
We do find, however, that the $J/\psi$ and $\Upsilon$ peaks stay in place, and that their windows defined in Table~\ref{tab:classes} lose at most 6\% of their pairs for $|\alpha| \le 3$.
We also find that the fit degrades for $|\alpha| \ge 1$ and splits the broadening between the width and
the resolution, so the fitted width in Fig.~\ref{fig:app_ladder_blocks} grows less than the peak.

\newpage
\subsection{Per-pair predictions}
\label{app:per_pair}
In Sec.~\ref{sec:steering_shape} we only compared the prediction of Eq.~\eqref{eq:prediction_of_steering}, which uses the average gradient of the class, with the fitted pole of the $Z$ boson.
Here, we shall take a step forward, and do this for each muon pair in Eq.~\eqref{eq:pair_steering}.

\begin{figure}[t]
\centering
\includegraphics[width=\textwidth]{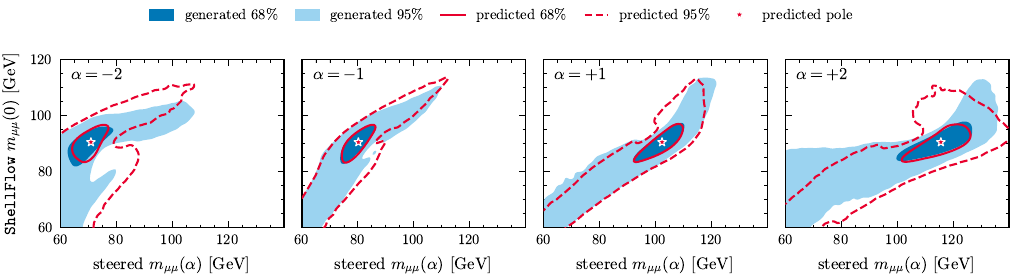}
\caption{Steered generation compared to the baseline \texttt{ShellFlow} generation for different values of $\alpha = -2, -1, +1, +2$.
The filled regions show the output of the steered model, and the red contour lines show the predicted steering of Eq.~\eqref{eq:pair_steering}.
The star is the result from Eq.~\eqref{eq:prediction_of_steering}.}
\label{fig:app_per_pair_contours}
\end{figure}

Figure~\ref{fig:app_per_pair_contours} compares the steered invariant mass of each pair with the prediction of
Eq.~\eqref{eq:pair_steering}.
The 68\% regions agree for all four values of $\alpha$.
The 95\% regions separate on the flanks of the $Z$, more at $|\alpha| = 2$.

From these predictions, we repeat the same process we did in the main text. 
We take the $Z$ pair candidates used for the steered generation in Sec.~\ref{sec:experiments}, and make an individual fit for each pair of the shape
\begin{equation}
    m_{\mu\mu,p}(\alpha) = m_{\mu\mu,p}(0)\,\exp\left(\alpha\,\mathcal{S}_p\right)\,,
\end{equation}
where $m_{\mu\mu,p}(0)$ is the baseline generation, $m_{\mu\mu,p}(\alpha)$ is the result after steering, and $\mathcal{S}_p$ is the free parameter.
We do it for the values $-3 \leq \alpha \leq 3$.
The exponential form describes each pair well (median $R^2 = 0.994$).
The fitted $\mathcal{S}_p$ and the computed $\sum_k \mathbf{G}_p(t_k)^\top\,\boldsymbol{\xi}_Z(t_k)$ differ by a median of 5\%.
Without a fit, Eq.~\eqref{eq:pair_steering} predicts the steered mass of each pair with a median error of 3\% of the pair's mass shift at $|\alpha| = 1$, and 7\% at $|\alpha| = 2$.
The exponential form fits worst near the window edges.
Within 4~GeV of them about 30\% of the pairs have $R^2 < 0.5$, against 2\% of the pairs between 89 and 95~GeV.

\subsection{\texorpdfstring{$J/\psi$}{J/psi} and \texorpdfstring{$\Upsilon$}{Upsilon} steering vectors}
\label{app:quarkonia}

\begin{figure}[t]
\centering
\includegraphics[width = \textwidth]{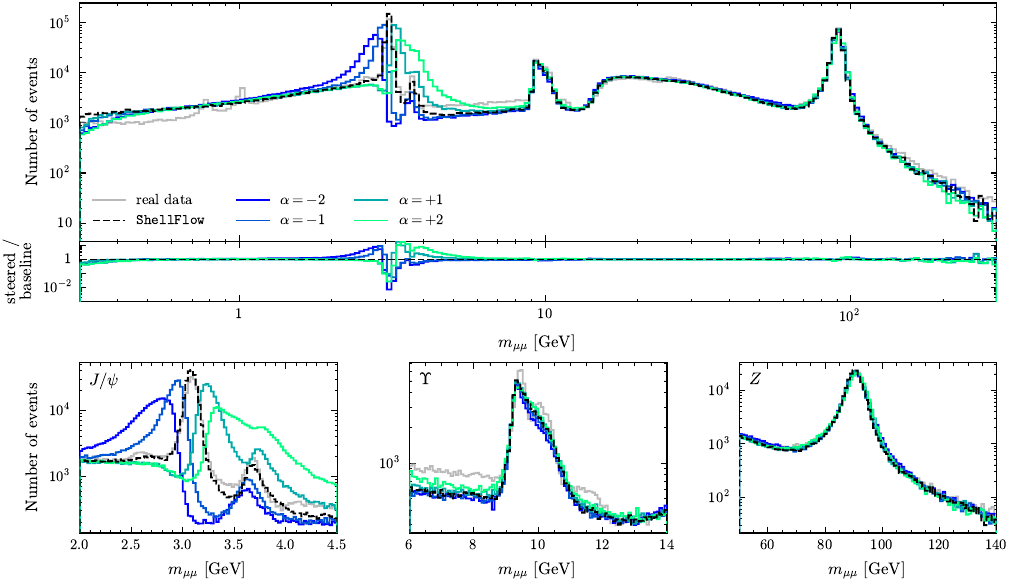}
\caption{The steering of the $J/\psi$ mass, as in Fig.~\ref{fig:spectrum}: dimuon invariant mass of one million muon pairs, compared with real data (grey), baseline generation of \texttt{ShellFlow} (black, dashed), and the results from adding the $J/\psi$ steering vector at block 1 with $\alpha = \pm 1, \pm 2$ (blue to green).}
\label{fig:app_spectrum_jpsi}
\end{figure}

\begin{figure}[t]
\centering
\includegraphics[width = \textwidth]{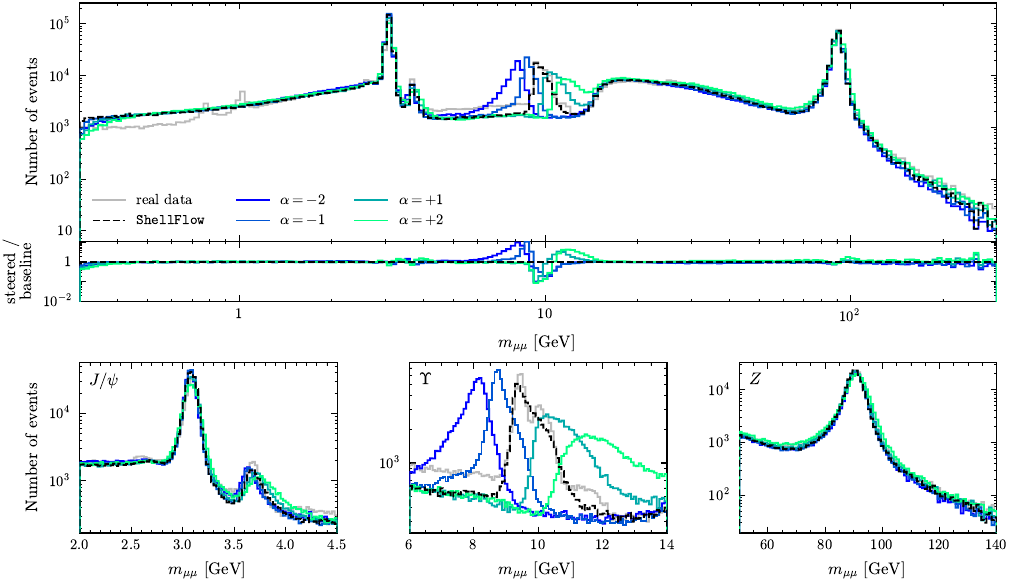}
\caption{The same as Fig.~\ref{fig:app_spectrum_jpsi} for the $\Upsilon$ steering vector.}
\label{fig:app_spectrum_upsilon}
\end{figure}

We build steering vectors for the $J/\psi$ and $\Upsilon$ classes of Table~\ref{tab:classes} in the same way as for
the $Z$.
We add them at block $L = 1$, with the same values of $\alpha$, the same 1,000,000 muon pairs, and the same three seeds
as in Sec.~\ref{sec:experiments}.
Figures~\ref{fig:app_spectrum_jpsi} and~\ref{fig:app_spectrum_upsilon} show the dimuon invariant mass for
$\alpha = \pm 1, \pm 2$.

We do not fit a lineshape to the quarkonia peaks, because it would require a careful understanding of the production channels.
Quarkonium production is still not fully understood. There is no single theoretical model that describes all measurements, in particular at low transverse momentum~\citep{lansberg-2020-quarkonia, brambilla-2025-nrqcd}.
Moreover, the $\Upsilon$ class holds three resonances that the model cannot separate (Table~\ref{tab:classes}).

\begin{wraptable}{r}{0.3\textwidth}
\caption{$\sum_k S_{\mathcal{C}}(t_k)$ of each class at block $L = 1$.}
\label{tab:class_response}
\centering
\small
\begin{tabular}{lc}
\toprule
Class $\mathcal{C}$ & $\sum_k S_{\mathcal{C}}(t_k)$ \\
\midrule
$Z$        & 0.122 \\
$J/\psi$   & 0.063 \\
$\Upsilon$ & 0.123 \\
\bottomrule
\end{tabular}
\end{wraptable}
The $J/\psi$ steering vector moves the $J/\psi$ peak from 3.08~GeV in the baseline generation to between 2.72 and
3.52~GeV for $|\alpha| \le 2$.
The $\psi(2S)$, which belongs to the rest class, moves with it. However, for positive $\alpha$ it becomes distorted.
For the same $\alpha$, the $J/\psi$ moves by about half the fraction of its mass that the $Z$ moves under the $Z$
steering vector.

The $\Upsilon$ steering vector moves the $\Upsilon$ peak from 9.68~GeV to between 7.95 and 12.81~GeV for
$|\alpha| \le 2$.
$\Upsilon$ also moves by about the same fraction of its mass as the $Z$ steering vector moves the $Z$.
The effect is not symmetric in $\alpha$.
For negative $\alpha$, the peak keeps its width down to $\alpha = -1$, and its width doubles at $\alpha = -2$.

Both quarkonium steering vectors move their own peak and leave the rest of the invariant mass spectrum alone.
Both broaden their own peak faster than the $Z$ steering vector broadens the $Z$.

\clearpage
\subsection{The steering vectors over flow time}
\label{app:time}

Here we look at the steering vectors of the $Z$, $J/\psi$, and $\Upsilon$ classes of Table~\ref{tab:classes} at each
of the 200 Euler steps, at block $L = 1$.

\begin{wrapfigure}{r}{0.47\textwidth}
\centering
\includegraphics[width = 0.45\textwidth]{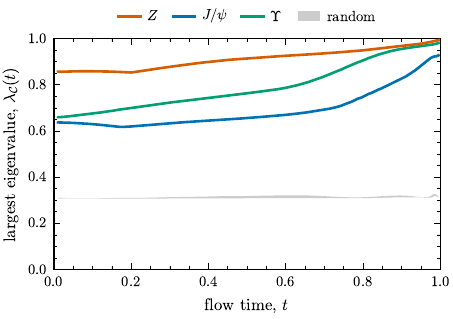}
\end{wrapfigure}
\paragraph{Separation of the classes.}
The figure on the right shows the largest eigenvalue $\lambda_{\mathcal{C}}(t)$ of Eq.~\eqref{eq:eigen} for each
class against flow time, as Fig.~\ref{fig:quality_steering}c does for the $Z$ at every block.
To know how large the values of $\lambda$ are, we build three random classes,
drawn from the pairs outside the $Z$ window.
We find the steering vector of each random class against all other pairs, with the same weighting as a resonance
against the other three classes.
Their largest eigenvalues form the grey band, at about 0.3 for the whole flow.
All three classes lie far above this band at every step, so a random group of pairs does not reach the separation of the classes.
The $Z$ is the most separable of the three at every step, and it is above 0.9 from $t \simeq 0.4$.
The $\Upsilon$ reaches 0.9 only at $t \simeq 0.8$, and the $J/\psi$ only in the last steps of the flow.
All three classes become more separable as the flow approaches $t = 1$.

\begin{wrapfigure}[15]{r}{0.47\textwidth}
\centering
\includegraphics[width = 0.45\textwidth]{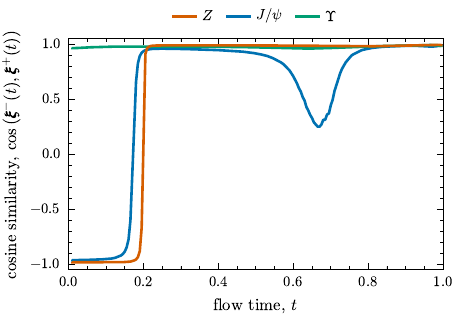}
\end{wrapfigure}
\paragraph{The two muons.}
In the main text we obtained two steering vectors, one to be added to the muon tokens, 
$\boldsymbol{\xi}^-$, and another to be added to the antimuon token $\boldsymbol{\xi}^+$.
The figure on the right shows the cosine similarity of both vectors.
At almost every step, the two parts are the same vector up to a sign. 
Two random vectors with 128 entries have a cosine similarity of about $\pm 1/\sqrt{128} \simeq \pm 0.09$.
For the $Z$ and the $J/\psi$, they point in opposite directions before $t \simeq 0.2$, and in the same direction after
it.
For the $\Upsilon$, they point in the same direction at every step.
Every change of sign falls where the two largest eigenvalues of Eq.~\eqref{eq:eigen} cross.
There, the steering vector changes from an eigenvector that moves the two muons in opposite directions to one that
moves them in the same direction.
The dip of the $J/\psi$ near $t \simeq 0.7$ is where its two largest eigenvalues come close without crossing.

\begin{wrapfigure}[14]{r}{0.47\textwidth}
\centering
\includegraphics[width = 0.45\textwidth]{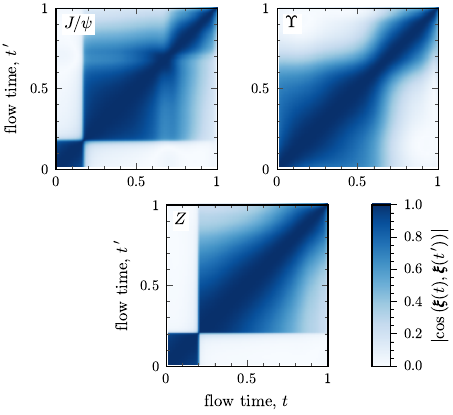}
\end{wrapfigure}
\paragraph{Rotation of the steering vectors.}
The figure on the right shows, for each class, the absolute cosine similarity between its steering vectors at two
flow times $t$ and $t^\prime$.
A steering vector that did not change with flow time would fill the map with ones.
Instead, all three rotate.
The $Z$ map splits into two blocks at $t \simeq 0.2$, where the parts on the two muons change sign: the steering
vectors before and after it are orthogonal.
After $t \simeq 0.2$, the $Z$ steering vector keeps turning slowly.
The $J/\psi$ splits at the same place, and it turns again near $t \simeq 0.7$, where its two parts dip.
The $\Upsilon$ turns smoothly, without a sharp change, and its early and late steering vectors are also orthogonal.
No steering vector stays within $|\cos| = 0.9$ of its final direction before the last two steps.
A single steering vector for all of the flow would not follow this rotation, which is why we use one for each Euler
step.

\section{Lineshape fit of the \texorpdfstring{$Z$}{Z} peak}
\label{app:lineshapefit}

We build a lineshape of the $Z$ boson to quantify the movement of the $Z$ pole. We do not claim this is a precision measurement of the $Z$ boson.
Every $Z$ pole, width, and resolution in this paper comes from one fit of one function to a histogram of the dimuon mass.
The same fit is applied to the real data, to the baseline generation, and to the generation at each $\alpha$, so the numbers
before and after an edit are directly comparable. Fits of this kind are standard at hadron colliders, for the mass
and width of the $Z$~\citep{abe-1989-measurement,aaij-2025-measurement} and for the calibration of the scale and
resolution of the muon momentum on the $Z$ peak~\citep{bodek-2012-extracting,cms-2018-performance,atlas-2023-studies}.
This appendix derives the function in four steps, the tree-level
cross section of the partons, the cross section of the protons, the detector resolution, and the final-state
radiation, states the order of each step, and then describes the fit.

At leading order a muon pair is produced by the annihilation of a quark and an antiquark through a photon or a $Z$
boson~\citep{drell-1970-massive},
\begin{equation}
q\,\bar q\;\to\;\gamma^*/Z\;\to\;\mu^-\,\mu^+,
\qquad \hat s=(p_q+p_{\bar q})^2=(p_-+p_+)^2=m^2 ,
\label{eq:process}
\end{equation}
with $m = m_{\mu\mu}$ the invariant mass of the pair. A fermion $f$ of charge $Q_f$ and weak isospin $T^3_f$ couples to the photon
with strength $eQ_f$ (where $e$ is electric charge) and to the $Z$ with the vector and axial couplings
\begin{equation}
v_f = T^3_f-2Q_f\,s_W^2,\qquad a_f=T^3_f
\label{eq:couplings}
\end{equation}
(where $c_W\equiv\cos\theta_W$, $s_W\equiv\sin\theta_W$, and $\theta_W$ is the Weinberg angle). With the photon
propagator taken out in front, the $Z$ enters through the ratio of the two propagators,
\begin{equation}
\chi(\hat s)=\frac{1}{4s_W^2c_W^2}\;\frac{\hat s}{\hat s-M^2+iM\Gamma},
\label{eq:chi}
\end{equation}
a Breit--Wigner resonance of mass $M$ and width $\Gamma$. The squared average amplitude is:
\begin{equation}
\begin{split}
\overline{\lvert\mathcal T\rvert^2}&=\frac{e^4}{3}
\Bigl\{\overbrace{\bigl[Q_q^2Q_\mu^2+2Q_qQ_\mu v_qv_\mu\operatorname{Re}\chi(\hat s)
+(v_q^2+a_q^2)(v_\mu^2+a_\mu^2)\,\lvert\chi(\hat s)\rvert^2\bigr]}^{\equiv A_0(\hat s)}\bigl(1+\cos^2\theta_\mathrm{CS}\bigr)\\
&\qquad+\underbrace{\bigl[4Q_qQ_\mu a_qa_\mu\operatorname{Re}\chi(\hat s)+8v_qa_qv_\mu a_\mu\lvert\chi(\hat s)\rvert^2\bigr]}_{\equiv A_1(\hat s)}\cos\theta_\mathrm{CS}\Bigr\}\\
&=\frac{e^4}{3}\Bigl[A_0(\hat s)\bigl(1+\cos^2\theta_\mathrm{CS}\bigr)+A_1(\hat s)\cos\theta_\mathrm{CS}\Bigr],
\end{split}
\label{eq:angular}
\end{equation}
where $\theta_\mathrm{CS}$ is the angle between $\mu^-$ and the quark in the rest frame of the pair, which at this order is the Collins--Soper angle. This leaves the tree-level cross section, with $\alpha_{\mathrm{em}}$ the electromagnetic coupling, as:
\begin{equation}
\hat\sigma_{q\bar q\to\mu^+\mu^-}(\hat s)=\frac{4\pi\alpha_{\mathrm{em}}^2}{9\,\hat s}\,A_0(\hat s).
\label{eq:sigmahat}
\end{equation}

The cross section of two protons at squared energy $s$, here $\sqrt s=13$~TeV, is the cross section of the partons
folded with the parton
distributions $f_q$ of the proton. This is the factorisation of \citet{drell-1970-massive}, established in QCD
by \citet{bodwin-1985-factorization} and \citet{collins-1989-factorization}:
\begin{equation}
\sigma=\sum_q\int_0^1\!\mathrm{d}x_1\int_0^1\!\mathrm{d}x_2\,
\bigl[f_q(x_1)f_{\bar q}(x_2)+f_{\bar q}(x_1)f_q(x_2)\bigr]\;\hat\sigma_{q\bar q\to\mu^+\mu^-}(x_1x_2s).
\label{eq:factorisation}
\end{equation}
The differential cross section in any variable follows by inserting one under the integral, written as the integral
of a delta function of that variable. For the mass of the pair,
$1=\int\mathrm{d}m^2\,\delta(m^2-x_1x_2s)$, and the integral over $x_2$ gives
\begin{equation}
\frac{\mathrm{d}\sigma}{\mathrm{d}m^2}=\frac1s\sum_q\hat\sigma_{q\bar q\to\mu^+\mu^-}(m^2)\,
\mathcal{L}_q\Bigl(\frac{m^2}{s}\Bigr),
\qquad
\mathcal{L}_q(\tau)=\int_\tau^1\frac{\mathrm{d}x}{x}\,
\Bigl[f_q(x)\,f_{\bar q}\Bigl(\frac{\tau}{x}\Bigr)+f_{\bar q}(x)\,f_q\Bigl(\frac{\tau}{x}\Bigr)\Bigr],
\label{eq:luminosity}
\end{equation}
with $\mathcal{L}_q$ the parton luminosity of flavour $q$.
\begin{figure}[t]
\centering
\includegraphics[width = \textwidth]{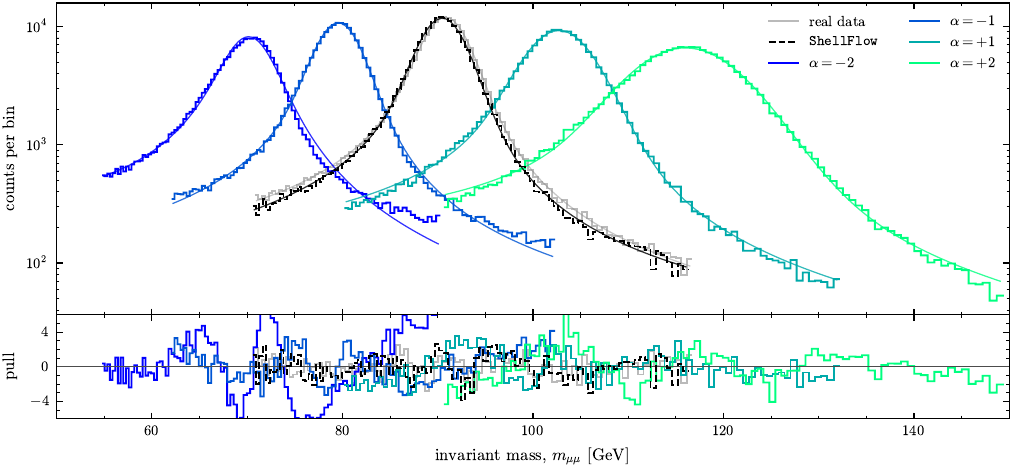}
\caption{Lineshape fits of the $Z$ peak. Real data (grey), baseline generation of \texttt{ShellFlow} (black,
dashed), and the generation steered at block~1 with $\alpha = -2, -1, +1, +2$ (blue to green), one seed. Each
sample is drawn in the 100 bins of its own fit window, with the fitted lineshape as a thin line. Bottom: pulls,
$(n_i - \nu_i)/\sqrt{\nu_i}$.}
\label{fig:app_fit_overlay}
\end{figure}

We compute the integral in the rapidity $Y$ of the pair, with $x = (m/\sqrt s)\,e^{Y}$ and $\mathrm{d}x/x = \mathrm{d}Y$,
over $|Y| < \log(\sqrt s/m)$. In terms of $\log m$,
\begin{equation}
\frac{\mathrm{d}\sigma}{\mathrm{d}\log m}=\frac{2}{s}\sum_q m^2\,\hat\sigma_{q\bar q\to\mu^+\mu^-}(m^2)\;
\mathcal{L}_q\Bigl(\frac{m^2}{s}\Bigr).
\label{eq:hadronic}
\end{equation}
This spectrum has been measured at the LHC for values of $m$ from GeV to TeV
energies~\citep{atlas-2014-measurement,atlas-2016-measurement,cms-2019-measurement}.

We take the luminosity of each quark type at the mass of the pair from the NNPDF4.0 distributions at
next-to-leading order~\citep{ball-2022-path}, with the scale equal to $m$, for two quark types, up-type and
down-type. We multiply it by the acceptance of the muon selection, $|\eta| < 2.5$ and $p_T > 7$~GeV, for pairs
without transverse momentum. A free power $(M/m)^{\rho}$ absorbs what this leading-order luminosity does not
describe: the higher orders, the transverse momentum of the pair, and the remaining backgrounds.
We define the number of pairs as the cross section times a normalisation constant $\mathcal{Y}$,
\begin{equation}
\frac{\mathrm{d}N_{\mathrm{Born}}}{\mathrm{d}\log m} \;=\;
\mathcal{Y}\,\Bigl(\frac{M}{m}\Bigr)^{\rho}\,\sum_q \frac{\mathcal{L}_q(m^2/s)}{\sum_{q'}\mathcal{L}_{q'}(M^2/s)}\;
m^2\,\hat\sigma_{q\bar q\to\mu^+\mu^-}(m^2),
\label{eq:born}
\end{equation}
where $\mathcal{L}_q$ now includes the acceptance. The constant $\mathcal{Y}$ includes the
integrated luminosity of the collider, the efficiency of the trigger and of the selection, and the constant factors
of Eq.~\eqref{eq:hadronic}. It is a free parameter, and its only use is to make the fit match the total number of pairs.

ATLAS~\citep{atlas-2008-atlas} measures the momentum of a muon with a resolution between one and a few
percent~\citep{atlas-2016-muon,atlas-2023-studies}. We describe this resolution by a
Gaussian in $\log m$ of standard deviation $\sigma_{\mathrm{det}}/M$,
the same over the fit window,
\begin{equation}
G(\mathfrak{l})=\frac{1}{\sqrt{2\pi}\,\sigma_{\mathrm{det}}/M}\,\exp\Bigl[-\frac{\mathfrak{l}^2}{2\,(\sigma_{\mathrm{det}}/M)^2}\Bigr],
\label{eq:resolution}
\end{equation}
cut at five standard deviations.

To account for final-state radiation, we introduce the radiator function $F$~\citep{montagna-1998-precision},
\begin{equation}
F(y) = 2b\,e^{-2y}\bigl(1-e^{-2y}\bigr)^{b-1},
\label{eq:fsr}
\end{equation}
where a pair with invariant mass $m'$ before radiation has invariant mass $m'e^{-y}$ after it, so $y \ge 0$. $b$ is defined as
\begin{equation}
b=\frac{2\alpha_{\mathrm{em}}}{\pi}\Bigl(\log\frac{M^2}{m_\mu^2}-1\Bigr),
\label{eq:exponent}
\end{equation}
evaluated at the fitted pole $M$, with $m_\mu$ the muon mass.
We model the $Z$ lineshape as
\begin{equation}
\frac{\mathrm{d}N}{\mathrm{d}\log m}=\int_0^{y_{\mathrm{cut}}}\!\mathrm{d}y\int\!\mathrm{d}\mathfrak{l}\;
\frac{\mathrm{d}N_{\mathrm{Born}}}{\mathrm{d}\log m}(\mathfrak{l})\;F(y)\;G\bigl(\log m-\mathfrak{l}+y\bigr),
\label{eq:observed}
\end{equation}
where $\mathfrak{l}$ is the logarithm of the invariant mass before radiation, and $\log m$ is the value after resolution. We set $y_{\mathrm{cut}} = 0.8$. This cutoff only amounts to a loss of ${\sim}1\%$ of the muon pairs.

\subsection{Details on the fit}
\label{app:fitdetails}
\begin{figure}[t]
\centering
\includegraphics[width = \textwidth]{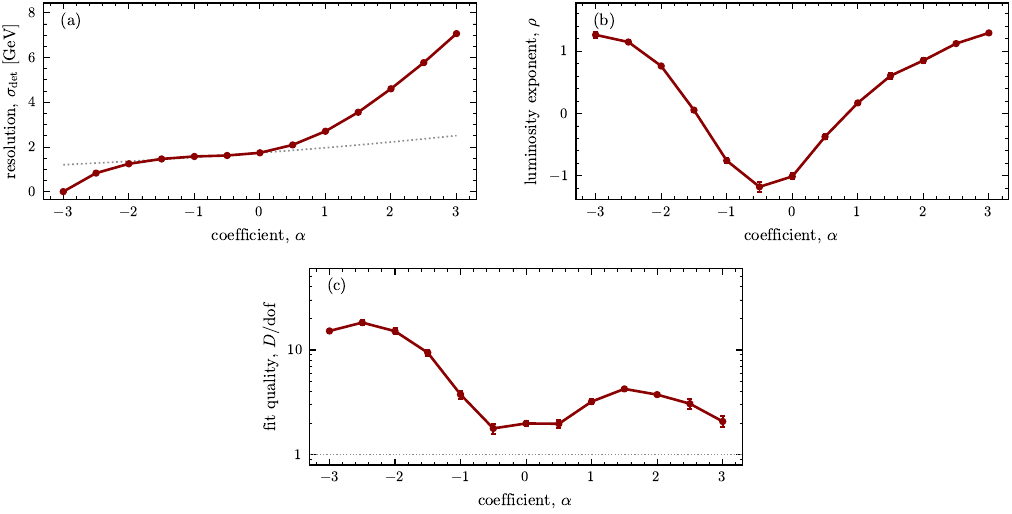}
\caption{The remaining parameters of the lineshape fit against $\alpha$, for the $Z$ steering vector at block~1:
(a) detector resolution $\sigma_{\mathrm{det}}$, (b) luminosity exponent $\rho$, and (c) deviance per degree of freedom.
Mean over three seeds, with their standard deviation as bars. Dotted in (a): the baseline value times
$\exp\left(\alpha\sum_{k}\bar{\mathbf{G}}_{Z}(t_k)^\top\boldsymbol{\xi}_{Z}(t_k)\right)$, a pure shift of
$\log m_{\mu\mu}$, as in Fig.~\ref{fig:ladder}. At $\alpha = -3$ the fit sets $\sigma_{\mathrm{det}}$ to zero and
puts all of the broadening into $\Gamma$.}
\label{fig:app_fit_vs_alpha}
\end{figure}
We fit Eq.~\eqref{eq:observed} to the counts of muon pairs in 100 bins of equal width in $\log m$, within
$\pm 0.25$ of the peak, about 71 to 117~GeV for the baseline generation. We first take as the peak the median
of $\log m$ over the pairs whose baseline mass lies in the $Z$ window of Table~\ref{tab:classes}, fit, then
centre the window on the fitted pole and fit again. Every pair of a sample enters the histogram, whatever its
mass before the steering.

The expected count $\nu_i$ in bin $i$ is the integral of Eq.~\eqref{eq:observed} over the bin, a function of
the five free parameters $\boldsymbol{\theta} = (M, \Gamma, \sigma_{\mathrm{det}}, \rho, \mathcal{Y})$. The observed counts $n_i$ are
Poisson distributed, and we minimise the deviance~\citep{baker-1984-clarification},
\begin{equation}
D(\boldsymbol{\theta}) = 2\sum_i \left[\nu_i(\boldsymbol{\theta}) - n_i + n_i \log\frac{n_i}{\nu_i(\boldsymbol{\theta})}\right],
\label{eq:deviance}
\end{equation}
which is $-2$ times the logarithm of the ratio of the Poisson likelihood of the fit to that of a perfect fit.
For large counts, $D$ follows a $\chi^2$ distribution with $100 - 5 = 95$ degrees of freedom, and we quote
$D$ per degree of freedom as the quality of the fit. The uncertainty of each parameter is where $D$ rises by
one above its minimum, $\Delta D = 1$, with the other parameters free. The Weinberg angle is fixed at
$\sin^2\theta_W = 0.23122$, and $b$ follows the fitted pole through Eq.~\eqref{eq:exponent}.

We evaluate the lineshape on a grid in $\log m$ from 25 to 320~GeV with a step of $10^{-3}$, and compute both
integrals of Eq.~\eqref{eq:observed} as discrete convolutions on this grid. We minimise $D$ with a bounded
trust-region method, from four starting points for $\Gamma$ and $\sigma_{\mathrm{det}}$, and keep the smallest $D$.
The pole is bounded to the fit window, $\Gamma$ and $\sigma_{\mathrm{det}}$ to between $10^{-4}$ and 0.35 times $M$,
and $\rho$ to $[-12, 15]$. The fit of the baseline starts at $\Gamma = 0.019\,M$, $\sigma_{\mathrm{det}} = 0.017\,M$,
and $\rho = 2.5$, with $M$ at the peak; each steered sample starts from the baseline fit of its seed, with $M$ at its
own peak.

\paragraph{Results.}
Table~\ref{tab:fitresults} gives the fits of the real data, the baseline generation, and the generation
steered with $-2 \le \alpha \le 2$ at block~1, and Fig.~\ref{fig:app_fit_overlay} shows the fitted lineshapes
against the counts. The statistical uncertainty of the pole is about 10~MeV,
far below the shifts of the steering, and the three generations with different seeds agree within it.
$\Gamma$ and $\sigma_{\mathrm{det}}$ have a correlation of $-0.8$ in every fit: the fit can trade the width of the
Breit--Wigner against the width of the resolution, which is why we do not read the growth of $\Gamma$ with
$|\alpha|$ as a change of the physical width alone. The fit describes the real data and the baseline well,
and degrades as $|\alpha|$ grows, most for negative $\alpha$, where the steered peak is no longer a
Breit--Wigner folded with a Gaussian (Fig.~\ref{fig:app_fit_vs_alpha}c). Fig.~\ref{fig:app_fit_vs_alpha} also
shows the remaining parameters of the fit against $\alpha$: the resolution grows faster than a pure shift of
$\log m_{\mu\mu}$ for $\alpha > 0$ and falls for $\alpha < 0$, while $\rho$ absorbs the change of the continuum on
both sides.

\begin{table}[t]
\caption{Lineshape fits of the $Z$ peak. Values are the mean over three seeds; uncertainties are statistical,
from $\Delta D = 1$. $D/\mathrm{dof}$ is the deviance per degree of freedom (95). PDG: world average of the
Particle Data Group~\citep{pdg-2026-review}.}
\label{tab:fitresults}
\centering
\def\dotrule{\noalign{\vskip 2pt\hbox to \columnwidth{\leaders\hbox to 3pt{\hss.\hss}\hfill}\vskip 1pt}}
\begin{tabular*}{\columnwidth}{@{\extracolsep{\fill}}lccccc@{}}
\toprule
Sample & $M$ [GeV] & $\Gamma$ [GeV] & $\sigma_{\mathrm{det}}$ [GeV] & $\rho$ & $D/\mathrm{dof}$\\
\midrule
PDG                 & $91.1879 \pm 0.0020$ & $2.4955 \pm 0.0023$ & -- & -- & --\\
\dotrule
Real data           & $90.979 \pm 0.009$  & $2.87 \pm 0.03$ & $1.60 \pm 0.02$ & $-0.32 \pm 0.06$ & 1.3\\
Baseline generation & $90.573 \pm 0.009$  & $2.41 \pm 0.02$ & $1.74 \pm 0.02$ & $-1.01 \pm 0.07$ & 2.0\\
\dotrule
$\alpha = -2$       & $70.299 \pm 0.011$  & $4.37 \pm 0.03$ & $1.25 \pm 0.03$ & $0.76 \pm 0.05$  & 15.1\\
$\alpha = -1.5$     & $74.769 \pm 0.010$  & $3.38 \pm 0.03$ & $1.47 \pm 0.02$ & $0.06 \pm 0.05$  & 9.4\\
$\alpha = -1$       & $79.677 \pm 0.009$  & $2.71 \pm 0.02$ & $1.58 \pm 0.02$ & $-0.76 \pm 0.06$ & 3.8\\
$\alpha = -0.5$     & $84.935 \pm 0.008$  & $2.41 \pm 0.02$ & $1.62 \pm 0.01$ & $-1.18 \pm 0.07$ & 1.8\\
$\alpha = +0.5$     & $96.634 \pm 0.010$  & $2.64 \pm 0.03$ & $2.10 \pm 0.02$ & $-0.37 \pm 0.07$ & 2.0\\
$\alpha = +1$       & $102.949 \pm 0.013$ & $3.00 \pm 0.03$ & $2.71 \pm 0.02$ & $0.17 \pm 0.07$  & 3.2\\
$\alpha = +1.5$     & $109.569 \pm 0.017$ & $3.42 \pm 0.04$ & $3.55 \pm 0.02$ & $0.61 \pm 0.07$  & 4.3\\
$\alpha = +2$       & $116.403 \pm 0.022$ & $3.89 \pm 0.05$ & $4.60 \pm 0.03$ & $0.85 \pm 0.08$  & 3.8\\
\bottomrule
\end{tabular*}
\end{table}

\end{document}